\documentclass[11pt,a4paper]{amsart}
\usepackage[margin=2.2cm]{geometry}
\usepackage{amsmath, amssymb, amsthm}
\usepackage{enumerate}
\usepackage{mathrsfs}
\usepackage{bbm}

\usepackage[round, comma]{natbib}
\numberwithin{equation}{section}

\newcommand{\norm}[1]{\left\lVert#1\right\rVert}
\newcommand{\abs}[1]{\left\lvert#1\right\rvert}

\newcommand{\bF}{\mathbb F}

\newcommand{\bN}{\mathbb N}

\newcommand{\bR}{\mathbb R}

\newcommand{\cC}{\mathcal C}

\newcommand{\cE}{\mathcal E}
\newcommand{\cF}{\mathcal F}
\newcommand{\cG}{\mathcal G}
\newcommand{\cH}{\mathcal H}

\newcommand{\cM}{\mathcal M}

\newcommand{\cP}{\mathcal P}
\newcommand{\cQ}{\mathcal Q}

\newcommand{\sF}{\mathscr F}
\newcommand{\sG}{\mathscr G}
\newcommand{\sH}{\mathscr H}

\newcommand{\sL}{\mathscr L}

\newcommand{\sP}{\mathscr P}
\newcommand{\sQ}{\mathscr Q}

\newcommand{\ind}{\mathbbm 1}

\DeclareMathOperator*{\essinf}{ess\,inf}
\DeclareMathOperator*{\esssup}{ess\,sup}

\DeclareMathOperator*{\pol}{Pol}

\DeclareMathOperator*{\pred}{Pred}
\newcommand{\rmG}{\textrm{G}}

\renewcommand{\cF}{\sF}
\renewcommand{\cG}{\sG}
\renewcommand{\cP}{\sP}
\renewcommand{\cQ}{\sQ}
\renewcommand{\cH}{\sH}
\usepackage[dvipsnames]{xcolor}

\usepackage[colorlinks=true, linkcolor = blue, citecolor = blue]{hyperref}

\begin{document}
	\theoremstyle{plain}
	
	\newtheorem{theorem}{Theorem}[section]
	\newtheorem{lemma}[theorem]{Lemma}
	\newtheorem{example}[theorem]{Example}
	\newtheorem{proposition}[theorem]{Proposition}
	\newtheorem{corollary}[theorem]{Corollary}
	\newtheorem{definition}[theorem]{Definition}
	\newtheorem{Ass}[theorem]{Assumption}
	\newtheorem{condition}[theorem]{Condition}
	\theoremstyle{definition}
	\newtheorem{remark}[theorem]{Remark}
	\newtheorem{SA}[theorem]{Standing Assumption}
	
	\makeatletter
	\newcommand{\mylabel}[2]{#2\def\@currentlabel{#2}\label{#1}}
	\makeatother
	
	\makeatletter
	\@namedef{subjclassname@2020}{%
		\textup{2020} Mathematics Subject Classification}
	\makeatother

	\title[Conditional second fundamental theorem]{A conditional version of the second fundamental theorem of asset pricing in discrete time}
	\author{Lars Niemann}
	\author{Thorsten Schmidt}
	\address{Department of Mathematical Stochastics, University of Freiburg, Germany}
	\email{Lars.Niemann@stochastik.uni-freiburg.de}
	\email{Thorsten.Schmidt@stochastik.uni-freiburg.de}
	\thanks{Financial support from the DFG in project SCHM 2160/13-1 is gratefully acknowledged. We thank two anonymous referees for their helpful comments and suggestions.}
	
	\date{\today}
	
	\begin{abstract}
		We consider a financial market in discrete time and study pricing and hedging conditional on the information available up to an arbitrary point in time. 
		In this conditional framework, we determine the structure of              arbitrage-free prices. Moreover, we characterize attainability and        market completeness. We derive a conditional version of the second fundamental theorem of asset pricing, which, surprisingly, is not      available up to now. 
		
		The main tool we use are time consistency properties of dynamic nonlinear expectations,  which we apply to the super- and subhedging prices. The results obtained extend existing results in the literature, where the conditional setting is  considered in most cases only on finite probability spaces.

	\end{abstract}

	\maketitle
	
	\section{Introduction}

	The mathematical analysis of financial markets starts with the remarkable 
	thesis of Bachelier, submitted to the Academy of Paris in 1900\nocite{Bachelier}. 
	More than half a century later, the search for a precise theory of option valuation
	was continued in  \cite{Samuelson1965} and encompassed with the observation that replication
	is  a key to pricing  in the famous works \cite{black1973pricing}, and \cite{merton1973theory} (honoured by the Nobel prize in Economics in 1997).

	The connection to martingales and martingale measures was started in the works \cite{harrison1979martingales} and \cite{harrison1981martingales},  which builds  the foundation for what we nowadays call the fundamental theorems of asset pricing. In continuous time, semimartingales turned out to play the central role and the fundamental theorems in this setting were established in a series of papers, see \cite{delbaen1994general, delbaen1998fundamental, delbaen2006mathematics} and references therein. 
	
	While the study of arbitrage in continuous time requires subtle arguments using semimartingale theory, the results simplify significantly in discrete time. In this realm, 
	the first fundamental theorem of asset pricing characterizes absence of arbitrage of a financial market by the existence of an equivalent martingale measure. It is the core of modern financial mathematics.
	
	The second fundamental theorem considers the more special case when a market is complete, i.e., when every European contingent claim can be replicated perfectly by a trading strategy, see  \cite{biagini2010second} for an overview and literature. It characterizes completeness by uniqueness of the prices for claims. This well-known result dates back to \cite{harrison1979martingales, harrison1981martingales, harrison1983stochastic}, although in the realm of continuous processes which excludes discrete time.
	In discrete time it was proven on a finite probability space in  \cite{taqqu1987analysis}, see also \cite{jacod1998local} and \cite{FS}.
	These results however consider only the initial time point $0$ and a conditional version of the second fundamental theorem is lacking. 
	
	This is the topic of the present article. 
	We derive a conditional formulation of the second fundamental theorem of asset pricing, based on the associated sub- and superhedging dualities.
	The proof of those dualities usually relies on the optional decomposition theorem, or on the simplification to a finite probability space. 
	Here, we will provide a self-contained proof of these dualities by using the theory of nonlinear expectations.
	
	Consider a discrete-time and arbitrage-free financial market with time horizon $T$. 
	The set of predictable trading strategies is denoted by \(\pred\).  
	A central quantity of interest is 
	the smallest conditional superhedging price
	\[
	\cE_t(H)     := \essinf \Big\{ H_t \in L^0(\Omega, \cF_t, P) : \exists \xi \in \pred : H_t +  
	\int_t^T \xi dX 
	\geq H\Big\} \,, \quad t \in \{0,\dots,T\},
	\]
	where the integral $\int_t^T \xi dX  = (\xi \cdot X)_T - (\xi \cdot X)_t$ is considered in discrete time.
	This expression gives rise to a nonlinear expectation, which turns out to be a central tool. 
	
	We will show that \((\cE_t)_{t \in \{0, \dots, T\}}\) is time-consistent, a key property to derive the associated conditional super- and subhedging dualities.
	More precisely, denote  by 
	\begin{align*}
		\pi^{\text{sup}}_t(H) & := \esssup \{ E_Q[H \mid \cF_t] : Q \in \cM_e^H\}
		, \quad \text{and} \\  
		\pi^{\text{inf}}_t(H) & := \essinf  \{ E_Q[H \mid \cF_t] : Q \in \cM_e^H\}
	\end{align*}
	the upper 
	and the lower 
	(conditional) bound of the no-arbitrage interval of a contingent claim $H$ with $\cM_e^H$ denoting the set of equivalent martingale measures under which $H$ is integrable. 
	We will establish in Proposition \ref{prop_price_boundaries}  that 
	\begin{equation} \label{eq: intro dualities}
		\cE_t(H) = \pi^{\text{sup}}_t(H)
	\end{equation}
	for every European contingent claim  $H \in L_+^0(\Omega, \cF_T, P)$. 
	If \(H\) can be superreplicated for a finite price, we obtain a similar expression for the lower bound \(\pi^{\text{inf}}_t(H)\).
	
	In the unconditional case, it is well-known  that the no-arbitrage interval collapses to a singleton if and only if the claim can be replicated 
	(see Section \ref{sec:attainability} for details).
	We generalize this result to the conditional setting in Theorem \ref{thm_attainable} where we show the following. 
	\begin{theorem} \label{thm: intro 1}
		Let $H \in L_+^0(\Omega, \cF_T, P)$ be a European contingent claim that       can be replicated for a finite price.
		Then, $H$ is attainable at time $t \in \{0,\dots,T\}$ if and only if 
		$$ \pi^{\text{inf}}_t(H) = \pi^{\text{sup}}_t(H).$$  
		If $H$ is not attainable at time $t$, then the prices $\pi^{\text{inf}}_t(H)$ and $\pi^{\text{sup}}_t(H)$ are not free of arbitrage.
	\end{theorem}
	In order to prove Theorem \ref{thm_attainable}, we argue on the basis of symmetry. In the realm of nonlinear expectations, symmetric elements play a major role, as they constitute the linear part of a nonlinear expectation. The symmetric elements for the nonlinear expectation \(\cE_t\) are precisely the claims attainable at time \(t\). Consequently, by means of the dualities provided along the lines of \eqref{eq: intro dualities}, attainability is equivalent to  \(E_Q[H | \cF_t]\) being constant over \(Q \in \cM_e^H\), i.e., equivalent to an unique arbitrage-free price.

	In Theorem \ref{FTAP2}, we provide  a conditional version of the second fundamental theorem of asset pricing by
	the following three (among five) equivalent conditions. 
	We denote by  \(Q \odot_t Q^* \)  the pasting of \(Q\) and \(Q^*\) at time \(t\).
	\begin{theorem} \label{thm: intro 2}
		For any $t \in \{0,\dots,T\}$, the following statements are equivalent:
		\begin{enumerate}[(i)]
			\item 
			every European contingent claim $H \in L_+^0(\Omega, \cF_T, P)$ is attainable at time $t$,
			\item every European contingent claim $H \in L_+^0(\Omega, \cF_T, P)$ has a unique price at time $t$,
			
			\item  for all \(Q, Q^* \in \cM_e\), it holds that  \(Q \odot_t Q^* = Q\). 
		\end{enumerate}
	\end{theorem}
	Since market completeness at time \(t > 0\) is a weaker property in comparison to completeness at time zero, the question arises if equivalence between completeness and uniqueness of the equivalent martingale measure prevails in the conditional case. 
	Part (iii) of Theorem \ref{thm: intro 2} answers this question: in a sense, there exists a unique equivalent martingale measure for the market from time $t$ on. More precisely, if a  pricing measure \(Q \in \cM_e\) is fixed up to time \(t\), then \(Q \odot_t \cM_e = \{Q\}\).
	
	The remarkable property that, in discrete time, the underlying probability space of a complete financial market is purely atomic (see, e.g., Theorem 5.37 in \cite{FS}) no longer persists in the conditional case. Yet, in a weaker sense, this property can be transported to markets complete at time \(t\): we will show in Theorem \ref{thm: atomic} that conditional on the atomic parts of \(\cF_t\),  the market from time \(t\) on remains complete and the underlying probability space is atomic. Intuitively, given the information \(\cF_t\), there exists a unique equivalent martingale measure for the market from time \(t\) on. Compared to the unconditional case, this gives an additional degree of freedom for modelling complete financial markets.

	\subsubsection*{Related literature}
	Nonlinear expectations and  dynamic risk measures have been studied in many places in the literature.
	Introduced  in \cite{peng2005nonlinear}, nonlinear expectations allow for a concise description of model uncertainty. This includes  $g$-expectation,  Brownian motion with uncertain drift and volatility   (see \cite{coquet2002filtration} and  \cite{peng2007g}).
	For applications of nonlinear expectations to discrete-time financial markets under uncertainty we refer to \cite{bartl2019}, \cite{blanchard2018}, \cite{bouchard2015arbitrage}, \cite{neufeld2018}, \cite{nutz2014superreplication}, \cite{nutz2016} and 
	references therein.
	
	Regarding  risk measures we refer to the seminal article \cite{artzner1999coherent}, to \cite{delbaen2002coherent} for unconditional risk measures, and to \cite{detlefsen2005conditional}, \cite{acciaio2011dynamic}, \cite{FS} for conditional and dynamic risk measures, amongst many others of course.

	The paper is structured as follows: in Section \ref{sec:supersubhedging} we study super- and subhedging and provide a new perspective on the optional decomposition theorem based on those. In Section \ref{sec:no-arbitrage-interval} we show that the set of arbitrage-free prices is indeed an interval and study further properties. 
	Section \ref{sec:FTAP} is the core of the paper and provides a detailed study of attainability and  the conditional version of the second fundamental theorem of asset pricing. In Appendix \ref{sec:nonlinear expectations} we introduce dynamic nonlinear expectations and derive related results on sensitivity and time consistency that are used throughout the paper.

	\section{Super- and subhedging}
	\label{sec:supersubhedging}
	
	Before dealing with the second fundamental theorem, we provide some results regarding
	super- and subhedging in a classical, finite-dimensional financial market in discrete time exploiting  time consistency of the associated nonlinear expectations.
	
	In this regard, fix a single probability measure $P$ and  consider a filtered probability space 
	$(\Omega, \cF, P, (\cF_t)_{t \in \{0,\dots,T\}})$. 
	Furthermore, assume that $\cF_0$ is trivial and $\cF_T = \cF$.
	We directly work on discounted prices which are described through the 
	$d$-dimensional discounted price process $X=(X_t)_{t \in \{0,\dots,T\}}$.
	Denote by $\pred$ the collection of $d$-dimensional predictable processes. 
	
	Throughout the paper, we identify a self-financing trading strategy with a predictable process  $\xi$. For such a process $\xi$ and  $t \in \{1,\dots,T\}$,  
	gains from self-financing trading 
	in $X$ until $t$ are given by 
	$$ (\xi \cdot X)_t=\sum_{k=1}^t \xi_k (X_k-X_{k-1}) = \sum_{k=1}^t \xi_k \Delta X_k.  $$ 
	An \emph{arbitrage} is a trading strategy $\xi$ such that 
	$$
	(\xi\cdot X)_T \geq 0, \quad P((\xi\cdot X)_T > 0) > 0.
	$$
	Denote by $\cM_e =\cM_e(P)$ the set of equivalent martingale measures. 
	We assume that the market is free of arbitrage which is equivalent to $\cM_e \neq \emptyset$. 
	
	Arbitrage-free prices will be studied in more detail in Section~\ref{sec:arbitrage-free-prices}. In particular, as a consequence of the first fundamental theorem, the set of arbitrage-free prices of 
	a bounded European contingent claim $H \in L^\infty(\Omega, \cF_T, P)$ is given by its expectations under all equivalent martingale measures.
	
	We denote its upper bound by
	\begin{align}\label{def:cE'_t(H)}
		\bar{\cE}_t(H) & := \esssup \{ E_Q[H \mid \cF_t] : Q \in \cM_e \},\quad t \in \{0,\dots,T\}.
	\end{align}
	
	It is  well-known that $\bar{\cE}$ is equal to the smallest superhedging price, 
	\begin{align}\label{def:cE_t(H)}
		\cE_t(H)     & := \essinf \{ H_t \in L^0(\Omega, \cF_t, P) : \exists \xi \in \pred : H_t + \rmG_t(\xi) \geq H\} \,,
		\quad t \in \{0,\dots,T\},
	\end{align}
	where \(H \in L^\infty(\Omega, \cF_T, P)\) and $\rmG_t(\xi) := (\xi \cdot X)_T - (\xi \cdot X)_t$. 
	
	While $\cE_0 = \bar{\cE}_0$ is easy to show,  the case $t \geq 1$ requires more sophisticated methods.
	We provide a new method for proving this result by deriving the superhedging  duality directly from the extension result Lemma~\ref{lem_consistent_unique}. To this end, we will show in Theorem \ref{thm_superhedge_consistent} below, that $\cE$ is a time-consistent expectation. As the same holds true for $\bar{\cE}$ (see Remark \ref{rem_pasting} below), Lemma \ref{lem_consistent_unique} gives access to the conditional superhedging duality $\cE_t = \bar{\cE}_t$ for $t \in \{1, \dots, T\}$.
	
	It is worth to recall the classical way to establish the superhedging duality. Usually, one verifies the $\cM_e$-supermartingale property of $\bar{\cE}$. 
	Then, one relies on the optional decomposition to show the difficult inequality $ \bar{\cE} \geq \cE$. We refer to \cite[Corollary 7.18]{FS} and \cite[Theorem 1, p.~514]{shiryaev} for more details.
	In case of a finite probability space, the proof simplifies as in \cite[Theorem 2.4.4]{delbaen2006mathematics}.
	
	In the following, we will frequently use techniques based on conditional nonlinear expectations. For the convenience
	of the reader, we collected the most important results for our analysis in Appendix \ref{sec:nonlinear expectations}. We refer to this place 
	for standard notions like sensitivity, time consistency, etc.
	As for processes, we will also need a notation for the dynamic conditional expectation. In this regard, we will denote $\cE(H)$ for the process $(\cE_t(H))_{t \in \{0,\dots,T\}}$ and use a similar notation for the other dynamic nonlinear expectations.

	\subsection{Superhedging prices}
	At time $t < T$, an $\cF_t$-measurable random variable is a superhedging price for the European contingent claim $H$ due at time $T$, if there exists a self-financing trading strategy $\xi$ which provides a terminal wealth greater than $H$, i.e.,
	$$ 
	H_t + \rmG_t(\xi) \geq H. 
	$$
	If equality can be achieved, i.e., when $H = H_t + \rmG_t(\xi)$,
	the claim $H$ can be perfectly hedged. Then it is called  \emph{attainable} (replicable) at time $t$.  
	
	We begin by developing some auxiliary results.
	The following lemma shows that for a $\cF_t$-measurable set $A$, the process $\ind_A \xi$ is again a self-financing trading strategy from $t$ on. We shall use this observation frequently in the following.
	
	\begin{lemma} \label{lem: eta}
		Fix \(t \in \{0,\dots,T\}\).
		Consider \(\xi \in \pred\) and  \(A \in \cF_t\). Then, there exists  \(\eta \in \pred\) such that 
		\[
		\rmG_t(\eta) = \ind_A \rmG_t(\xi).
		\]
	\end{lemma}
	\begin{proof}
		We define the process \(\eta\) by
		\[
		\eta_k := \begin{cases} 0 ,& \text{for } k \leq t,\\
			\ind_A \xi_k,& \text{for } k \geq t+1.
		\end{cases}
		\]
		Then, $\eta \in \pred$ has the desired property.
	\end{proof}

	Next, we show that the smallest superhedging defined in  \eqref{def:cE_t(H)} is a sublinear dynamic expectation which is also sensitive\footnote{The mapping $\cE_t:L^\infty(\Omega, \cF_T, P) \to L^\infty(\Omega, \cF_t, P)$ is called \(\cF_t\)-\emph{conditional expectation}, if it is monotone and preserves \(\cF_t\)-measurable functions. If, additionally, $\cE_t(H+H') \le  \cE_t(H)+ \cE_t(H')$ and $\cE_t(X_t\, H) = X_t\, \cE_t(H)$ for all $X \in L^\infty(\Omega, \cF_t, P)$ with $X_t \ge 0$ and all $H,H' \in L^\infty(\Omega, \cF_T, P)$, it is called $\cF_t$-\emph{sublinear}.
		It is called \emph{sensitive}, if for every $H \in  L^\infty(\Omega, \cF_T, P)$ with $H \ge 0$ and $\cE_t(H)=0$ it follows that $H=0$. If those properties hold for all $t \in \{0,\dots,T\}$, we say that the dynamic nonlinear expectation $\cE$ has this property.
		We refer to appendix \ref{sec:nonlinear expectations} for all details.
	} on the space of bounded random variables. Note that sensitivity is implied by absence of arbitrage, see Remark \ref{rem:sensitivity} below for a detailed discussion.

	\begin{lemma}
		\label{lem_sublinear_expectation}
		The superhedging price $\cE$ is a sensitive, sublinear dynamic expectation on the space $L^\infty(\Omega, \cF_T, P)$. 
		In particular, \(\cE_t(0) = 0\) for every \(t \in \{0, \dots, T\}\).
	\end{lemma}
	\begin{proof}
		First, we show that $\cE_t(H)$ is bounded for $H \in L^\infty(\Omega, \cF_T, P)$. The inequality $ \cE_t(H) \leq \norm{H}$ follows by definition.
		If $H_t \in L^0(\Omega, \cF_t, P)$ is a superhedging price, choose $\xi \in \pred$ with
		$ H_t + \rmG_t(\xi) \geq H$.
		Consider the set $A := \{ H_t < - \norm{H} \} \in \cF_t$.
		Thanks to Lemma \ref{lem: eta}, there exists $\eta \in \pred$ with
		\[
		\rmG_t(\eta) = \ind_A \rmG_t(\xi) \geq \ind_A(H-H_t) \geq 0,
		\]
		and therefore $P(A) = 0$ by absence of arbitrage.
		We conclude that $ - \norm{H} \leq \cE_t(H) \leq \norm{H}$.
		
		Next, we show that \(\cE\) preserves measurable random variables, i.e., that
		\(\cE_t(H) = H\) for every \(H \in L^\infty(\Omega, \cF_t, P)\).
		If \(H \in L^\infty(\Omega, \cF_t, P)\), note that \(\cE_t(H) \leq H\) by definition of \(\cE_t\). Define \( B := \{ \cE_t(H) < H \} \in \cF_t\), and let \(H_t\) be a superhedging price for \(H\) with corresponding strategy \(\xi \in \pred\).
		Using Lemma \ref{lem: eta} again, we obtain a strategy \(\eta \in \pred\) with
		\[
		\rmG_t(\eta) = \ind_B \rmG_t(\xi) \geq \ind_B ( H - H_t) \geq \ind_B ( H - \cE_t(H)) \geq 0,
		\]
		and thus $P(B) = 0$ by absence of arbitrage. This implies \(\cE_t(H) = H\).
		Moreover, one readily checks the remaining properties of a sublinear expectation. 
		To finish the proof, note that sensitivity of $\cE_t$ follows from the no-arbitrage assumption.
	\end{proof}

	\subsection{Symmetry}
	The concept of  \emph{symmetric} random variables plays a major role in the analysis of perfect hedges.
	A random variable $H$ is called symmetric with respect to $\cE$ if
	$$ \cE(H) = - \cE(-H).$$
	This motivates another  dynamic expectation $\cE^*$ associated to $\cE$ given by
	\begin{align}\label{defin:E*} \cE^*_t(H) := - \cE_t(-H). \end{align}
	The \emph{symmetric elements} of \(\cE_t\), are now those random variables  with $
	\cE^*_t(H) =  \cE_t(H).$
	Lemma 2.2 in \cite{cohen2011sublinear} underlines the importance of symmetry:  for every sublinear conditional expectation \(\cE_t\), it holds that
	\begin{equation} \label{eq: symmetry}
		\cE_t(H^1 + H^2) = \cE_t(H^1) + \cE_t(H^2)
	\end{equation}
	for arbitrary \(H^1\) and  \emph{symmetric} \(H^2\).
	In this sense, 
	the symmetric elements constitute the linear part of a nonlinear expectation.

	\subsection{European contingent claims}
	Up to now, we treated only bounded random variables, which excludes for example European calls. More generally, we are interested in \emph{European contingent claims} (claims for short) which are simply nonnegative random variables.
	To include them in our study, we extend the domain of the sublinear expectations \(\cE_t\) and \(\bar{\cE}_t\) 
	to $L^0(\Omega, \cF_T, P)$. 
	
	Note that, in contrast to set of claims \(L^0_+(\Omega, \cF_T, P)\), the  space $L^0(\Omega, \cF_T, P)$ is symmetric in the sense that 
	\[
	L^0(\Omega, \cF_T, P) = - L^0(\Omega, \cF_T, P),
	\]
	which allows us to utilize symmetry as introduced above.
	The extension of $\cE$ is rather straightforward:
	we define for \(H \in L^0(\Omega, \cF_T, P)\)
	\[\cE_t(H) := \essinf \{ H_t \in L^0(\Omega, \cF_t, P) : \exists \xi \in \pred : H_t + \rmG_t(\xi) \geq H\} \,,
	\quad t \in \{0,\dots,T\}.
	\]
	Note that the properties stated in Lemma \ref{lem_sublinear_expectation} transfer to this larger space.
	
	However, defined on $L^0(\Omega, \cF_T, P)$, the nonlinear expectation $\cE_t$ has, in general,  no longer finite values. 
	We will prove in Proposition \ref{prop: values} below that it takes values  in $(- \infty, +\infty]$. Note that this is a consequence of  no-arbitrage.
	Likewise, the associated expectation $\cE_t^*$ takes values in $[-\infty, \infty)$. In particular, if $H \in L^0(\Omega, \cF_T, P)$ is symmetric with respect $\cE_t$, then
	$\cE_t(H)$ is finite-valued.
	
	\bigskip

	A first step in this direction is to show that minimum of two superhedging prices is again a superhedging price, i.e.,~the set of superhedging prices is directed downwards.
	\begin{lemma} \label{lem: directed downward}
		Let \(H \in L^0(\Omega, \cF_T, P)\). Then, the set 
		\[
		M := \{ H_t \in L^0(\Omega, \cF_t, P) : \exists \xi \in \pred : H_t + \rmG_t(\xi) \geq H \}
		\]
		of superhedging prices is directed downwards.
	\end{lemma}
	\begin{proof}
		Let \(H^1_t, H^2_t \in M\) and pick strategies \(\xi^1, \xi^2 \in \pred\) such that 
		\[
		H^1_t + \rmG_t(\xi^1) \geq H, \quad H^2_t + \rmG_t(\xi^2) \geq H.
		\]
		We define \(A := \{H^1_t \leq H^2_t\} \in \cF_t\). Thanks to Lemma \ref{lem: eta}, there exist \(\eta^1, \eta^2 \in \pred\) such that 
		\[
		\rmG_t(\eta^1) = \ind_A \rmG_t(\xi^1), \quad \rmG_t(\eta^2) = \ind_{A^c} \rmG_t(\xi^2).
		\]
		Set \(\eta := \eta^1 + \eta^2 \in \pred\). Then, 
		\[
		(H^1_t \wedge H^2_t) + \rmG_t(\eta) = (H^1_t \wedge H^2_t) + \ind_A \rmG_t(\xi^1) + \ind_{A^c} \rmG_t(\xi^2) \geq H,
		\]
		and we conclude that \( H^1_t \wedge H^2_t \in M\).
	\end{proof}

	\begin{proposition} \label{prop: values}
		Let \(H \in L^0(\Omega, \cF_T, P)\). Then, 
		\(\cE_t(H)\) takes values in \((-\infty, + \infty]\).
	\end{proposition}
	\begin{proof}
		Set \(A := \{\cE_t(H) = - \infty \} \in \cF_t\). We will show that \(A\) has probability zero.
		We exploit that  the set $M := \{ H_t \in L^0(\Omega, \cF_t, P) : \exists \xi \in \pred : H_t + \rmG_t(\xi) \geq H \} $ of superhedging prices is directed downwards, which was shown in Lemma \ref{lem: directed downward}.
		Then, the essential supremum can be approximated by a decreasing sequence. More precisely, 
		\cite[Theorem A.33]{FS} grants existence of  a sequence $(H_t^n)_{n \in \bN} \subseteq M$ with $H_t^n \downarrow \cE_t(H)$ a.s. 
		
		By construction, we may pick for each $n \in \bN$ a strategy \(\xi^n \in \pred\) such that
		\[
		\rmG_t(\xi^n) \geq H - H^n_t \geq (H-H^n_t) \wedge 1.
		\]
		For each \(n \in \bN\), Lemma \ref{lem: eta} grants the existence of a process \(\eta^n \in \pred\)
		with
		\begin{equation} \label{eq: pf prop values}
			\rmG_t(\eta^n) = \ind_A \rmG_t(\xi^n) \geq \ind_A \left( (H-H^n_t) \wedge 1 \right).
		\end{equation} 
		Define, for \(n \in \bN\), the random variable \(g^n := \ind_A \left( (H-H^n_t) \wedge 1 \right)\). 
		Then, \eqref{eq: pf prop values} shows that \(g^n \in \cC\), where
		\[ \cC:= \{ \rmG_t(\xi) - U : \xi \in \pred,~ U \in L^0_+(\Omega, \cF_T, P) \}.
		\]
		As the cone \(\cC\) is closed due to the no-arbitrage assumption, see \cite[Theorem 6.9.2]{delbaen2006mathematics} or \cite[Theorem 1]{stricker}, it follows that 
		\(\lim_n g_n = \ind_A \in \cC\).
		Using again the no-arbitrage assumption implies \(P(A) = 0\).
	\end{proof}
	
	\begin{remark}[Symmetry of gains processes]
		\label{rem_orthogonality}
		For every $H \in L^0(\Omega, \cF_T, P)$, and every $\xi \in \pred$ one has the orthogonality
		\begin{align} \label{eq:orthogonality}
			\cE_t(H + \rmG_t(\xi)) = \cE_t(H) .
		\end{align}
		This follows from the observation that the set of superhedging prices for $H$ coincides with the set of superhedging prices for $H + \rmG_t(\xi)$. In particular, \eqref{eq:orthogonality} implies for every $\xi \in \pred$
		\begin{align*} 
			\cE_t(\rmG_t(\xi)) = \cE_t(0) = 0,
		\end{align*}
		and likewise
		\begin{align*} 
			\cE^*_t(\rmG_t(\xi)) = \cE^*_t(0) = 0.
		\end{align*}
		We conclude that \(\rmG_t(\xi)\) is symmetric with respect to \(\cE_t\).
		Therefore, \eqref{eq:orthogonality} emerges as a special case of Equation \eqref{eq: symmetry}.
		More generally, we will show in Lemma \ref{lem_symmetric_replicable} below that attainable random variables are symmetric.
	\end{remark}	
	
	The following result shows that $\cE_t(H)$ is indeed a superhedging price. The idea is to construct a monotone approximating sequence of superhedging prices and then use closedness of the cone of super-replicable claims (which is implied by absence of arbitrage).

	\bigskip
	
	\begin{lemma}
		\label{lem_superhedging_price}
		Let  $t \in \{0,\dots,T\}$. Then,
		$\cE_t(H)$ is a superhedging price for $H\in L^0(\Omega, \cF_T, P)$.
	\end{lemma}
	\begin{proof}
		As in the proof of Proposition \ref{prop: values}, 
		we approximate the essential supremum by a decreasing sequence,
		i.e., we obtain a sequence of superhedging prices $(H_t^n)_{n \in \bN}$ with $H_t^n \downarrow \cE_t(H)$ a.s. 
		Then, for each $n \in \bN$,
		\begin{equation} \label{eq: pf superhedging}
			H = H^n_t + \rmG_t(\xi^n) - U^n, 
		\end{equation}
		for some $\xi^n \in \pred$ and $U^n \in L^0_+(\Omega, \cF_T, P)$.
		Recall that \(\cE_t(H)\) takes values in \((-\infty, \infty]\) by Proposition \ref{prop: values}. On the set \(A:= \{\cE_t(H) = \infty\} \in \cF_t\), \(\cE_t(H)\)
		is a superhedging price for \(H\).
		Using Lemma \ref{lem: eta}, we may pick a sequence \((\eta^n)_{n \in \bN} \subseteq \pred \) with 
		\( \rmG_t(\eta^n) = \ind_{A^c} \rmG_t(\xi^n)\) for each \(n \in \bN\).
		Then, it follows from \eqref{eq: pf superhedging} that 
		\[
		(H-H^n_t)\ind_{A^c} = \rmG_t(\eta^n) - U^n\ind_{A^c}.
		\]
		As \((H-H^n_t)\ind_{A^c}\) converges to the finite-valued random variable \((H-\cE_t(H))\ind_{A^c}\) for \(n \to \infty\), and since the cone $\{ \rmG_t(\xi) - U : \xi \in \pred,~ U \in L^0_+(\Omega, \cF_T, P) \}$
		is closed due to the no-arbitrage assumption,  see \cite[Theorem 6.9.2]{delbaen2006mathematics} or \cite[Theorem 1]{stricker}, we may pick \(\eta \in \pred\)
		with
		\[
		H\ind_{A^c} \leq \cE_t(H)\ind_{A^c} + \rmG_t(\eta).
		\]
		Consequently, we obtain
		\[
		H = H \ind_A + H \ind_{A^c} \leq \cE_t(H) \ind_A + \cE_t(H)\ind_{A^c} + \rmG_t(\eta) = \cE_t(H) + \rmG_t(\eta),    
		\]
		and the claim follows. 
	\end{proof}

	The previous result grants a concrete representation of the process of superhedging prices $\cE(H)$. As a first result, we will build a connection to appropriately generalized martingale properties of dynamic nonlinear expectations in the following lemma. 
	In this regard, we call a process  $Y$ an \(\cM_e\)-(super-/sub-)martingale, if it is a \(Q\)-(super-/sub-)martingale for every \(Q \in \cM_e\).
	Further on, 
	Corollary \ref{cor_replicable_unique_price}  will show  that for any $t \in \{0,\dots,T\}$, the process
	$(\cE_s(H))_{s\geq t}$ is an $\cM_e$-martingale, if and only if $H$ is attainable at time $t$ and therefore allows to characterize  the martingale property in terms of attainability.
	
	\begin{lemma}
		\label{lem_supermartingale}
		For every bounded European contingent claim $H \in L^\infty(\Omega, \cF_T, P)$, the process of superhedging prices $\cE(H)$ is an $\cM_e$-supermartingale. 
	\end{lemma}
	\begin{proof}
		Fix  $t \in \{0,...,T-1\}$. 
		Using Lemma \ref{lem_superhedging_price}, we may choose $\xi \in \pred$ with
		\begin{align*} \cE_t(H) + G_t(\xi) &\geq H.  \end{align*}
		Next, we apply $\cE_{t+1}$ to this inequality. 
		Note that 
		\[
		G_t(\xi) = G_{t+1}(\xi) + \xi_{t+1}\Delta X_{t+1} \text{ and }   \cE_{t+1}(\cE_t(H))=\cE_t(H).
		\]
		From Equation \eqref{eq:orthogonality} it follows that 
		\[
		\cE_{t+1}(\rmG_t(\xi)) = \cE_{t+1}(\xi_{t+1} \Delta X_{t+1} + \rmG_{t+1}(\xi)) = \cE_{t+1}(\xi_{t+1} \Delta X_{t+1}) = \xi_{t+1} \Delta X_{t+1}.
		\]
		Hence, 
		\begin{align} \label{eq:temp 632}
			\cE_t(H) + \xi_{t+1}\Delta X_{t+1} \geq \cE_{t+1}(H).
		\end{align} 
		For the $\xi$ chosen above, $\xi_{t+1}\Delta X_{t+1}$ is bounded from below. Thus, its conditional expectation under $Q$ vanishes for all $Q \in \cM_e$ by \cite[Theorem 2]{jacod1998local}. We obtain
		the inequality 
		\begin{align*} 
			\cE_t(H) \geq E_Q[\cE_{t+1}(H) \mid \cF_t],  \end{align*}
		for each $Q \in \cM_e$ and each $t \in \{0,...,T-1\}$, which implies the claim.
	\end{proof}
	
	\subsection{Extending the upper bound of the no-arbitrage interval}
	The extension of \(\bar{\cE}_t\) defined in \eqref{def:cE'_t(H)} from bounded random variables to the set 
	\(L^0(\Omega, \cF_T, P)\) is done as follows:
	for any random variable \(H\) and every probability measure \(Q\) on \((\Omega, \cF_T)\), we define the (generalized) conditional expectation by
	\[
	E_Q[H | \cF_t] := E_Q[H^+  | \cF_t] - E_Q[H^-  | \cF_t],
	\] 
	with the convention \( \infty - \infty := -\infty\). 
	Then, we may safely define $\bar\cE$ for any $H \in L^0(\Omega, \cF_T, P)$  by
	\[
	\bar{\cE}_t(H) = \esssup \{ E_Q[H \mid \cF_t] : Q \in \cM_e \}.
	\]
	For such \(H \in L^0(\Omega, \cF_T, P)\), we deduce from Proposition \ref{prop_prices} below the existence of \(Q \in \cM_e\) with 
	\(E_Q[H^-] < \infty\). In particular, \(E_Q[H | \cF_t]\) is \((-\infty, \infty]\)-valued, and so is \(\bar{\cE}_t(H)\).

	Recall from \cite[Theorem 1.17]{HWY} that for every random variable \(H \in L^0(\Omega, \cF_T, P)\),  the conditional expectation
	\( E_Q[|H| | \cF_t]\) is finite-valued if and only if \(H\) is \(\sigma\)-integrable with respect to \(\cF_t\), i.e., if there exists a sequence \((A^n)_{n \in \bN} \subseteq \cF_t\) with \(\ind_{A^n} \uparrow 1\) and 
	\(H \ind_{A^n} \in L^1(\Omega, \cF_T, Q)\) for each \(n \in \bN\).
	We will show in Lemma \ref{lem: integrability} below that for every \(H \in L^0_+(\Omega, \cF_T, P)\) with \( \cE_t(H) < \infty \) \(P\)-a.s., and every \(\xi \in \pred\) such that
	\[
	\cE_t(H) + \rmG_t(\xi) \geq H,
	\]
	the random variable \(\rmG_t(\xi)\) is \(\sigma\)-integrable with respect to \(\cF_t\) under any \(Q \in \cM_e\).
	
	\subsection{Super- and subhedging dualities}
	
	In this section we establish the conditional superhedging duality 
	\[
	\cE_t = \bar{\cE}_t, \text{ for } t \in \{1, \dots, T\}.
	\]
	
	As a first step in this direction we show that  the smallest superhedging price $\cE$ defined in Equation \eqref{def:cE_t(H)} is  time-consistent, i.e.,~$\cE_s = \cE_s \circ \cE_t$ for all $s \le t$ (see Section \ref{sec:sensitivity and time consistency}).

	\begin{theorem}
		\label{thm_superhedge_consistent}
		The dynamic nonlinear expectation $\cE$ 
		is time-consistent on $L^\infty(\Omega, \cF_T, P)$.
	\end{theorem}
	\begin{proof}
		Time consistency is equivalent to  $\cE_{t} = \cE_{t} \circ \cE_{t+1}$ for all $t \in \{0,\dots,T-1\}$. We first show $\cE_{t} \le \cE_{t} \circ \cE_{t+1}$. In this regard, applying Lemma \ref{lem_superhedging_price} to the European contingent claims $\cE_{t+1}(H)$ and $H$ allows to choose strategies $\xi, \eta \in \pred$ such that
		\begin{align*}  \cE_t(\cE_{t+1}(H)) + \rmG_t(\xi) \geq \cE_{t+1}(H)   \end{align*}
		and
		$$ \cE_{t+1}(H) + \rmG_{t+1}(\eta) \geq H .$$
		Combining both inequalities yields
		$$ \cE_{t}(\cE_{t+1}(H)) + \rmG_t(\xi) + \rmG_{t+1}(\eta) \geq H. $$
		Let \(\lambda \in \pred\) be a strategy with \(\rmG_t(\lambda) = \rmG_t(\xi) + \rmG_{t+1}(\eta)\).
		Then, the claim $H$ can be super-replicated at time $t$ at price $\cE_{t}(\cE_{t+1}(H))$. As $\cE_t(H)$ is by definition the smallest superhedging  price for claim $H$ at time $t$, we obtain $$ \cE_{t}(\cE_{t+1}(H)) \geq \cE_t(H).$$  
		
		Concerning the reverse inequality, we obtain from Lemma \ref{lem_superhedging_price} the existence of $\theta \in \pred$, such that
		$$ H \leq \cE_{t}(H) + \rmG_{t}(\theta) .$$
		From Equation \eqref{eq:orthogonality} it follows that
		$$ \cE_{t+1}(\rmG_t(\theta)) = \cE_{t+1}(\theta_{t+1} \Delta X_{t+1} + \rmG_{t+1}(\theta)) = \cE_{t+1}(\theta_{t+1} \Delta X_{t+1}) = \theta_{t+1} \Delta X_{t+1}. $$
		This implies
		$$  \cE_{t+1}(H) \leq \cE_t(H) +  \theta_{t+1} \Delta X_{t+1}.  $$
		As before, pick \(\tilde{\lambda} \in \pred\) with \(\rmG_t(\tilde{\lambda}) = \theta_{t+1} \Delta X_{t+1}\) and  conclude that
		\begin{align*} 
			\cE_t(\cE_{t+1}(H)) &\leq \cE_t(H). \qedhere
		\end{align*}
	\end{proof}

	\begin{remark}[Time consistency of $\bar\cE$]
		\label{rem_pasting}
		Time consistency of $\bar{\cE}$ is related to the stability of $\cM_e$. 
		A set of equivalent probability measures $\cM$ is said to be \emph{stable} (under pasting),
		if for any two measures $Q^1, Q^2$, and every $t \in \{0, \dots, T\}$, the pasting  $ Q^1 \odot_t Q^2$ of $Q^1$ and $Q^2$ in $t$, defined by
		\begin{align} \label{def:pasting}
			Q^1 \odot_t Q^2 (A) := E_{Q^1}[ E_{Q^2}[ \ind_A \mid \cF_t]] \quad (A \in \cF_T)
		\end{align}
		is contained in $\cM$. In case of $\cM_e$, this is easily verified, see Proposition 6.42 in \cite{FS}, as for any $s, t \in \{0, \dots, T\}$ one has
		$$ E_{Q^1 \odot_t Q^2}[ \, \cdot \,  | \cF_s] = E_{Q^1}[ E_{Q^2}[ \, \cdot \,  | \cF_{s \vee t}] \mid \cF_s] \,,$$
		by \cite[Lemma 6.41]{FS}.
		Theorem 11.22 in \cite{FS} now shows that the expectation $\bar{\cE}$ is time-consistent.
		
	\end{remark}

	Since both $\cE$ and $\bar \cE$ are time-consistent, the uniqueness result in Lemma \ref{lem_consistent_unique} readily implies the superhedging duality for bounded claims.
	
	\begin{corollary}
		\label{cor_superhedging_duality}
		For every $t \in \{0,\dots,T\}$, and every $H \in L^\infty(\Omega, \cF_T, P)$, the superhedging duality 
		\begin{align}\label{def:suphedging_bounded}
			\cE_t(H) &= \bar{\cE}_t(H)
		\end{align}
		holds.
	\end{corollary}
	
	\begin{proof}
		Thanks to Lemma \ref{lem_sublinear_expectation}, \(\cE\) is a sublinear expectation. By construction, the same holds true for \(\bar{\cE}\). Notice that every subadditive expectation is translation-invariant.
		Hence, Proposition \ref{prop: trans} implies that both, \(\cE\) and \(\bar{\cE}\)
		are local on \(L^\infty(\Omega, \cF_T, P)\).
		In view of Theorem \ref{thm_superhedge_consistent} and Remark \ref{rem_pasting}, Lemma \ref{lem_consistent_unique} implies the claim.
		
	\end{proof}
	
	\begin{remark}\label{rem:sensitivity}
		Let us quickly discuss sensitivity of the superhedging price at time zero
		$$\cE_0(H) = \inf\{x \in \bR \colon \exists \xi \in \pred \colon x + (\xi \cdot X)_T \geq H\}, \quad H \in L^\infty(\Omega, \cF_T, P),  $$
		in more detail. It is easy to see that sensitivity of \(\cE_0\) is equivalent to absence of arbitrage in the financial market. 
		In view of Remark \ref{rem_sensitivity_polars}, this is in line with Corollary \ref{cor_superhedging_duality}. Indeed, consider the nonlinear expectation
		$$ \hat{\cE}_0(H) := \sup_{Q \in \cM_a} E_Q[H] ,\quad H \in L^\infty(\Omega, \cF_T, P), $$
		where $\cM_a$ denotes the set of martingale measures \emph{absolutely continuous} with respect to $P$.
		Assume, for the moment, that $\cM_a$ is nonempty, while $\cM_e$ is possibly empty.
		Then, by Remark \ref{rem_sensitivity_polars}, sensitivity of $\hat{\cE}_0$ is equivalent to $\cM_a \sim P$. In this case, the Halmos-Savage theorem implies that the set of equivalent martingale measure $\cM_e$ is nonempty.
	\end{remark}

	\begin{remark}\label{rem:3.8}
		It is worth to revisit how the equality $\cE_0 = \bar{\cE}_0$ is usually established, as it 
		can be formulated in terms of nonlinear expectations. To ease the argument, assume that the price process is (locally) bounded.
		Denote by $K_0 := \{ G_0(\xi) : \xi \in \pred \}$ the set of claims attainable at price 0,
		and by $C := (K_0 - L^0_+(\Omega, \cF_T, P)) \cap L^\infty(\Omega, \cF_T, P)$ the corresponding cone.
		By the very definition of $C$, a measure $Q \sim P$ is contained in            $\cM_e$ if and only if
		$E_Q[H] \leq 0$ for every $H \in C$.
		The Bipolar Theorem then implies that $H \in C$ if and only if
		$E_Q[H] \leq 0$ for every $Q \in \cM_e$.
		This can also be read as: the acceptance set $\{H \in L^\infty(\Omega, \cF_T, P) \colon \bar{\cE}_0 \leq 0 \}$        of $\bar{\cE}_0$ coincides with $C$.
		One readily verifies that by construction $C = \{H \in L^\infty(\Omega, \cF_T, P) \colon \cE_0 \leq 0\}$, and translation-invariance of both, $\cE_0$ and $\bar{\cE}_0$, implies $\cE_0 = \bar{\cE}_0$.
	\end{remark}
	
	\begin{corollary}
		\label{cor_subhedging_duality}
		For every $H \in L^\infty(\Omega, \cF_T, P)$, the subhedging duality
		\begin{align}\label{def:subhedging_bounded}
			\esssup \{ H_t \in L^0(\Omega, \cF_t, P) : \exists \xi \in \pred : H_t + \rmG_t(\xi) \leq H\}
			& = \essinf \{ E_Q[H \mid \cF_t] : Q \in \cM_e \} \,,
		\end{align}
		holds.
	\end{corollary}

	\begin{proof}
		This follows from Corollary \ref{cor_superhedging_duality}, as 
		\begin{align} \label{def:E*}
			\cE^*_t(H) = \esssup \{ H_t \in L^0(\Omega, \cF_t, P) : \exists \xi \in \pred : H_t + \rmG_t(\xi) \leq H \}
		\end{align}
		and 
		\begin{align*} 
			(\bar{\cE}_t)^*(H) & = \essinf \{ E_Q[H \mid \cF_t] : Q \in \cM_e \} \,. \qedhere
		\end{align*}
	\end{proof}

	\subsection{The extension to unbounded claims}
	Up to now, super- and subhedging dualities were only established for bounded European contingent claims.
	The next result shows continuity from below of the nonlinear conditional expectations $\cE$ and $\bar \cE$, which allows to extend the dualities to European contingent claims.
	
	A nonlinear conditional expectation $\cE_t$ is called \emph{continuous from below} on $L^0_+(\Omega, \cF_T, P)$, if for every sequence  $(H^n)_{n \in \bN} \subseteq L^0_+(\Omega, \cF_T, P)$ with $H^n \uparrow H \in L^0_+(\Omega, \cF_T, P)$ the equality
	$$ \cE_t(H) = \sup_n \cE_t(H^n) $$
	holds. If this property holds for all times, we  call $\cE$ continuous from below. 
	\begin{lemma}
		\label{lem_hedging_continuous}
		The expectations $\cE$ and $\bar{\cE}$ are continuous from below on $L^0_+(\Omega, \cF_T, P)$.
	\end{lemma}
	\begin{proof}
		Let $(H^n)_{n \in \bN} \subseteq L^0_+(\Omega, \cF_T, P)$ with $H^n \uparrow H \in L^0_+(\Omega,\cF_T, P)$.
		We start with $\bar{\cE}$. In this case, continuity from below is already entailed in the robust representation of $\bar{\cE}$. For convenience of the reader, we provide the details.
		By monotonicity, we have $\sup_n \bar{\cE}_t(H^n) \leq \bar{\cE}_t(H)$.
		For the other inequality, choose a sequence $(Q^m) \subseteq \cM_e$ with
		$$ \sup_m E_{Q^m}[H \mid \cF_t] = \esssup_{Q \in \cM_e} E_Q[H \mid \cF_t] \,.$$
		Monotone convergence then implies
		\begin{align*}
			\bar{\cE}_t(H) & = \sup_m E_{Q^m}[~\sup_n H^n \mid \cF_t] \\
			& = \sup_m \sup_n E_{Q^m}[H^n \mid \cF_t] \\
			& \leq \sup_n \bar{\cE}_t(H^n).
		\end{align*} 
		For \(\cE\), we have $\sup_n \cE_t(H^n) \leq \cE_t(H)$ by monotonicity.
		Regarding the converse inequality, we will show that  \(\sup_n \cE_t(H^n)\)
		is a superhedging price for \(H\). We argue similar as in the proof of Lemma \ref{lem_superhedging_price}.
		By Lemma \ref{lem_superhedging_price}, we may write for each $n \in \bN$,
		\begin{equation} \label{eq: pf continuous}
			H^n = \cE_t(H^n) + \rmG_t(\xi^n) - U^n, 
		\end{equation}
		for some $\xi^n \in \pred$ and $U^n \in L^0_+(\Omega, \cF_T, P)$.
		Recall that, for \(n \in \bN\), \(\cE_t(H^n)\) takes values in \((-\infty, \infty]\) by Proposition \ref{prop: values}.
		On the set \(A:= \{\sup_n \cE_t(H^n) = \infty\} \in \cF_t\), \(\sup_n \cE_t(H^n)\)
		is a superhedging price for \(H\).
		Using Lemma \ref{lem: eta}, we may pick a sequence \((\eta^n)_{n \in \bN} \subseteq \pred \) with 
		\( \rmG_t(\eta^n) = \ind_{A^c} \rmG_t(\xi^n)\) for each \(n \in \bN\).
		Then, it follows from \eqref{eq: pf continuous} that 
		\[
		(H^n-\cE_t(H^n))\ind_{A^c} = \rmG_t(\eta^n) - U^n\ind_{A^c}, \quad n \in \bN.
		\]
		As \( \left(H^n-\cE_t(H^n) \right)\ind_{A^c}\) converges to the finite-valued random variable \((H-\sup_n \cE_t(H^n))\ind_{A^c}\) for \(n \to \infty\), and since the cone $\{ \rmG_t(\xi) - U : \xi \in \pred,~ U \in L^0_+(\Omega, \cF_T, P) \}$
		is closed due to the no-arbitrage assumption,  see \cite[Theorem 6.9.2]{delbaen2006mathematics} or \cite[Theorem 1]{stricker}, we may pick \(\eta \in \pred\)
		with
		\[
		H\ind_{A^c} \leq \big(\sup_n \cE_t(H^n)\big)\ind_{A^c} + \rmG_t(\eta).
		\]
		Consequently, we obtain
		\[
		H = H \ind_A + H \ind_{A^c} \leq \big(\sup_n \cE_t(H^n)\big) \ind_A + \big(\sup_n \cE_t(H^n)\big)\ind_{A^c} + \rmG_t(\eta) = \sup_n \cE_t(H^n) + \rmG_t(\eta),    
		\]
		which shows that \(\sup_n \cE_t(H^n)\) is a superhedging price for \(H\).
	\end{proof}

	\begin{proposition}
		\label{prop_superhedging_claims}
		The superhedging duality \eqref{def:suphedging_bounded}, and time consistency of $\cE$ extends to claims, i.e., to $L^0_+(\Omega, \cF_T, P)$.
	\end{proposition}
	\begin{proof}
		In view of Corollary \ref{cor_superhedging_duality}, and Theorem \ref{thm_superhedge_consistent}	this follows from Lemma \ref{lem_hedging_continuous}.
	\end{proof}

	In the context of Remark \ref{rem:3.8}, 
	Proposition \ref{prop_superhedging_claims} allows us to state the following conditional
	version of the Bipolar relationship:
	a claim $H \in L^0_+(\Omega, \cF_T, P)$ can be super-replicated at time $t$ at price zero, if and only if $E_Q[H \mid \cF_t] \leq 0$ for every
	$Q \in \cM_e$.
	
	\begin{lemma}
		\label{lem_subhedging_price}
		For every claim $H \in L^0_+(\Omega, \cF_T, P)$, $\cE^*_t(H)$ defined in \eqref{def:E*} is a subhedging price for $H$.
	\end{lemma}
	\begin{proof}
		Similar to Lemma \ref{lem_superhedging_price}.
	\end{proof}
	\begin{proposition}
		\label{prop_subhedging_claims}
		Consider the claim $H\in L^0_+(\Omega, \cF_T, P)$. If $\cE_t(H)$ is          finite-valued, then the subhedging duality \eqref{def:subhedging_bounded} holds
		at time $t$.
	\end{proposition}
	\begin{proof}
		Pick $\xi \in \pred$ with
		$\cE_t(H) + \rmG_t(\xi) \geq H$, and consider the claim
		$\tilde{H} := \cE_t(H) + \rmG_t(\xi) - H$.
		Applying the superhedging duality in form of Proposition \ref{prop_superhedging_claims} to $\tilde{H}$ yields 
		$$ \cE_t(\cE_t(H) + \rmG_t(\xi) - H) = \bar{\cE}_t(\cE_t(H) + \rmG_t(\xi) - H) \,.$$
		Using translation in conjunction with Remark \ref{rem_orthogonality}, in particular Equation \eqref{eq:orthogonality}, gives
		$$ \cE_t(H) + \cE_t(-H) = \cE_t(H) + \bar{\cE}_t(-H) $$
		and the claim follows.
	\end{proof}

	We briefly collect the associated symmetric statements of Lemma \ref{lem_hedging_continuous}, Proposition \ref{prop_superhedging_claims} and Proposition \ref{prop_subhedging_claims}. We call a nonlinear conditional expectation $\cE_t$ \emph{continuous from above} on the set $L^0_-(\Omega, \cF_T, P)$, if for every sequence  $(H^n)_{n \in \bN} \subseteq L^0_-(\Omega, \cF_T, P)$ with $H^n \downarrow H \in L^0_-(\Omega, \cF_T, P)$ the equality
	$$ \cE_t(H) = \inf_n \cE_t(H^n) $$
	holds. If this property holds for all times, we  call $\cE$ continuous from above. 
	
	\begin{proposition}
		The following statements hold true:
		\begin{enumerate}[(i)]
			\item The expectations $\cE^*$ and $(\bar{\cE})^*$ are continuous from above on $L^0_-(\Omega, \cF_T, P)$.
			
			\item The subhedging duality \eqref{def:subhedging_bounded}, and time consistency of $(\cE^*_t)$ extend to $L^0_-(\Omega, \cF_T, P)$.
			
			\item Consider $H \in L^0_-(\Omega, \cF_T, P)$. If $\cE^*_t(H)$ is finite-valued, then the superhedging duality \eqref{def:suphedging_bounded} holds at time $t$.
		\end{enumerate}
	\end{proposition}

	We end this section with an integrability result.
	Here, the conditional setting fundamentally differs from the unconditional one. Indeed, a claim $H$ attainable at time zero for a finite price can always be replicated by a bounded initial investment since $\cF_0$ is trivial.
	This is no longer the case for $t>0$. Here, one has to pay a finite, but not necessarily bounded, price. In particular, the gains $\rmG_t(\xi) = H - H_t$ might be unbounded from below and expectations might no longer exist. Clearly, this relates to the observation that for arbitrary $\xi \in \pred$ the stochastic integral
	$((\xi \cdot X)_t)_{t \in \{0,\dots,T\}}$ is in general only a local martingale under $Q \in \cM_e$ due to lacking integrability.
	The link between integrability and the martingale property is discussed in  \cite{jacod1998local, meyer2006martingales} in detail.

	\begin{lemma} \label{lem: integrability}
		Consider a claim $H\in L^0_+(\Omega, \cF_T, P)$. If $\cE_t(H)$ is          finite-valued, then
		\begin{enumerate}
			\item[(i)] for every \(\xi \in \pred\) with \(\cE_t(H) + \rmG_t(\xi) \geq H\), it holds that
			\[
			E_Q[| \rmG_t(\xi)| |\cF_t] < \infty, \quad \text{and } E_Q[\rmG_t(\xi) |\cF_t] = 0
			\]
			for every \(Q \in \cM_e\), and            
			
			\item[(ii)] for every \(\eta \in \pred\) with \(\cE^*_t(H) + \rmG_t(\eta) \leq H\), it holds that 
			\[
			E_Q[| \rmG_t(\eta)| |\cF_t] < \infty, \quad \text{and } E_Q[\rmG_t(\eta) |\cF_t] = 0
			\]
			for every \(Q \in \cM_e\).
		\end{enumerate}
	\end{lemma}
	\begin{proof}
		Let \(Q \in \cM_e\) be arbitrary.
		We start with part \((i)\). Let \(\xi \in \pred\) with \(\cE_t(H) + \rmG_t(\xi) \geq H\). Set \(A^n := \{\cE_t(H) \leq n \} \in \cF_t\). As \(\cE_t(H)\) is finite-valued by assumption, we have \(\ind_{A^n} \uparrow 1\). Further, for each \(n \in \bN\), 
		\[
		\rmG_t(\xi) \ind_{A^n} \geq H \ind_{A^n} - \cE_t(H) \ind_{A^n} \geq - n,
		\]
		which shows that \(\rmG_t(\xi) \ind_{A^n}\) is bounded from below.
		Thanks to Lemma \ref{lem: eta}, for each \(n \in \bN\), there exists \(\tilde{\xi}^n \in \pred\)
		with \(\rmG_t(\tilde{\xi}^n) = \rmG_t(\xi)\ind_{A^n}\).
		Hence, \cite[Theorem 2]{jacod1998local} implies that \(\rmG_t(\xi)\ind_{A^n} \in L^1(\Omega, \cF_T, Q)\). We conclude that \( E_Q[|\rmG_t(\xi)| |\cF_t]\) is finite-valued. Using \cite[Theorem 1.21]{HWY}, we obtain
		\[
		E_Q[\rmG_t(\xi) | \cF_t] \ind_{A^n} = E_Q[ \rmG_t(\xi) \ind_{A^n} | \cF_t] = E_Q[\rmG_t(\tilde{\xi}^n) | \cF_t],
		\]
		for each \(n \in \bN\). Using again that \(\rmG_t(\tilde{\xi}^n)\) is bounded from below, a second application of \cite[Theorem 2]{jacod1998local} yields
		\(E_Q[\rmG_t(\tilde{\xi}^n) | \cF_t] = 0\) for every \(n \in \bN\). Therefore, \( E_Q[\rmG_t(\xi) | \cF_t] = 0\) as claimed.
		We proceed with part \((ii)\). Let \(\eta \in \pred\) with 
		\begin{equation} \label{eq: pf integrability}
			\cE^*_t(H) + \rmG_t(\eta) \leq H
		\end{equation}
		and consider the claim
		\[
		\tilde{H} := \cE_t(H) + \rmG_t(\xi) - H
		\]
		for any superhedging strategy \(\xi \in \pred\).
		Then, 
		\[
		\cE_t(\tilde{H}) = \cE_t(H) + \cE_t(-H) = \cE_t(H) - \cE^*_t(H)
		\]
		is finite-valued. From \eqref{eq: pf integrability}, we obtain
		\[
		\cE_t(\tilde{H}) + \rmG_t(\xi-\eta) \geq \tilde{H}.
		\]
		Using part \((i)\), we know that \(\rmG_t(\xi-\eta)\) and \(\rmG_t(\xi)\) are
		\(\sigma\)-integrable with respect to \(\cF_t\). Hence, \(\rmG_t(\eta)\) is \(\sigma\)-integrable with respect to \(\cF_t\) and
		\begin{align*}
			E_Q[\rmG_t(\eta) | \cF_t] & =  - E_Q[\rmG_t(\xi-\eta) | \cF_t] +  E_Q[\rmG_t(\xi) | \cF_t] = 0. \qedhere
		\end{align*}
	\end{proof}
	
	\subsection{The optional decomposition}
	While it is possible to derive time consistency of the superhedging prices $\cE$ using the optional decomposition, by taking advantage of the  $\cM_e$-supermartingale property of $\cE$ established in Lemma \ref{lem_supermartingale}, let us mention that the superhedging duality in Proposition \ref{prop_superhedging_claims} allows us to prove the optional decomposition with little effort. We record this in Theorem \ref{thm_optional_decomposition} below, whose proof is similar to the proof of \cite{delbaen2006mathematics}, Theorem 2.6.1.
	
	\begin{theorem} \label{thm_optional_decomposition}
		Let $V$ be a nonnegative $\cM_e$-supermartingale. Then, there exists an adapted, increasing process $C$ with $C_0 = 0$, and a strategy $\xi \in \pred$ such that
		$$ V_t = V_0 + (\xi \cdot X)_t - C_t \,,\quad t \in \{0,\dots,T\}. $$	
	\end{theorem}
	\begin{proof}
		For every $t \in \{1, \dots, T\}$, and every $Q \in \cM_e$ the inequality
		$$ E_Q[V_t \mid \cF_{t-1}] \leq V_{t-1} $$
		holds. Equivalently, this can be written in terms of the nonlinear expectation $\bar{\cE}$ as $\bar{\cE}_{t-1}(V_t) \leq V_{t-1}$.
		Applying the superhedging duality Proposition \ref{prop_superhedging_claims} to the claim
		$ V_t \in L^0_+(\Omega, \cF_t, P)$, we conclude that
		$$ \cE_{t-1}(\Delta V_t) \leq 0 .$$
		Hence, for $t \in \{1, \dots, T\}$ there exists a strategy $\xi^{(t)} \in \pred$ such that
		$$ \Delta V_t \leq \rmG_{t-1}(\xi^{(t)}) .$$
		Using the orthogonality relation as introduced in Remark \ref{rem_orthogonality}, in particular Equation \eqref{eq:orthogonality}, an application of $\cE_t$ yields
		$$ \Delta V_t \leq \xi^{(t)}_t \Delta X_t . $$
		Summing over $t \in \{0, \dots, T\}$, we obtain $\xi \in \pred$ such that
		$(\xi \cdot X) - V$ is increasing.
	\end{proof}
	
	In continuous time, the optional decomposition theorem, and the resulting construction of superhedging strategies were first obtained in \cite{el1995dynamic} in the setting of continuous processes. The extension to general locally bounded semimartingales was achieved in \cite{kramkov1996optional}. Later, the assumption of local boundedness was removed, see \cite{follmer1997optional, delbaen1999compactness} and references therein. For the statement in discrete time, we refer to \cite{follmer1997optional}, Theorem 2, and \cite{FS}, Theorem 7.5. The case of a finite probability space in discrete time is treated in \cite{delbaen2006mathematics}, Theorem 2.6.1.

	\section{The conditional no-arbitrage interval}
	\label{sec:no-arbitrage-interval}
	
	This section studies the conditional no-arbitrage interval and we begin with some
	results on  pricing conditional on  past information. 
	It seems to be worthwhile pointing out that in case of a finite probability space, conditional pricing can be reduced to the unconditional setting by exploiting the structure of the set of equivalent martingale measures.
	
	The no-arbitrage interval has been intensively studied in the unconditional case where $t=0$. We refer once more to \cite{FS}, Section 5 for a detailed treatment. In the following, we extend the notions and results therein to arbitrary $t \in \{0,\dots,T\}$.

	\subsection{Structure of arbitrage-free prices}\label{sec:arbitrage-free-prices}
	The main goal in computing arbitrage-free prices relying on  the fundamental theorem of asset pricing 
	is to obtain a price process for a new security which can be added to the market without violating absence of arbitrage.
	
	In this spirit, a random variable $\pi_t \in L^0(\Omega, \cF_t, P)$ is called \emph{arbitrage-free price}  at time $t$ of a European contingent claim $H\in L^0(\Omega, \cF_T, P)$  if there exists an adapted process
	$X^{d+1}$ such that $X^{d+1}_t = \pi_t$, $X^{d+1}_T = H$ and the  market $(X,X^{d+1})$ extended by $X^{d+1}$ is free of arbitrage. Denote by $\Pi_t(H)$ the collection of arbitrage-free prices.

	For every claim $H\in L^0_+(\Omega, \cF_T, P)$ we define
	$$ \cM_e^H := \{ Q \in \cM_e : H \in L^1(\Omega, \cF_T, Q) \} \,.$$
	
	Note that for an attainable claim $H$, the equality	$\cM^H_e = \cM_e$ holds.
	The fundamental theorem of asset pricing implies immediately that the set of arbitrage-free prices is given by expectations under the risk-neutral measures, which we state here for clarity.\footnote{The unconditional version of this result is, for example, given in Theorem 5.29 in \cite{FS}.}
	
	\begin{proposition}
		\label{prop_prices}
		For every claim $H\in L^0_+(\Omega, \cF_T, P)$,
		$$ \Pi_t(H) = \{ E_Q[H \mid \cF_t] : Q \in \cM_e^H \} \,, $$
		and the set of arbitrage-free prices is nonempty.
	\end{proposition}
	
	The following lemma shows that already when a risk-neutral conditional expectation has finite values, it is an arbitrage-free price.
	
	\begin{lemma}
		\label{lem_price_finite}
		Let $H\in L^0_+(\Omega, \cF_T, P)$ and $Q \in \cM_e$.
		If $E_Q[H \mid \cF_t]$ is finite-valued, then it is an arbitrage-free price.
	\end{lemma}
	\begin{proof}
		If $E_Q[H \mid \cF_t]$ is finite-valued, it is in $L^0_+(\Omega, \cF_t, P)$, and by Proposition \ref{prop_prices} there
		exists $\tilde{Q} \in \cM_e$ such that
		$E_Q[H \mid \cF_t]$ is integrable with respect to $\tilde{Q}$. We now paste $Q$ and $\tilde Q$.
		As mentioned in Remark \ref{rem_pasting}, the pasting $\tilde{Q} \odot_t Q$ of $\tilde{Q}$ with $Q$ in $\cF_t$  given by \eqref{def:pasting} is again an equivalent martingale measure. By construction, we even have  $\tilde{Q} \odot_t Q \in \cM_e^H$. Moreover, it follows that
		\begin{align*} 
			E_{\tilde{Q} \odot_t Q}[H \mid \cF_t] & = E_Q[H \mid \cF_t] \,. 
		\end{align*}
		This is an arbitrage-free price by Proposition \ref{prop_prices}.
	\end{proof}

	Denote the upper and the lower bound of the no-arbitrage set at time $t$ by
	\begin{align*}
		\pi^{\text{sup}}_t(H) & := \esssup \Pi_t(H), \quad \text{and}  \quad
		\pi^{\text{inf}}_t(H)  := \essinf  \Pi_t(H).
	\end{align*}	
	A priori, it is unclear if $\Pi_t(H)$ is indeed a (random) interval.
	While in the unconditional case this follows immediately, since \(\Pi_0(H)\) is the image of the convex set \(\cM_e\) under the map \(Q \mapsto E_Q[H]\) this is more subtle in the conditional case:
	consider a  \(\sigma\)-field \(\cG \subset \cF\) and 
	\(Q^1, Q^2 \in \cM_e\), \(0 \leq \lambda \leq 1\) together with \(Q := \lambda Q^1 + (1-\lambda) Q^2\).
	Observe that it is not necessarily true that 
	\[
	E_Q[H | \cG] = \lambda E_{Q^1}[H | \cG] + (1- \lambda) E_{Q^2}[H | \cG]
	\]
	holds and hence, convexity of $\Pi_t(H)$ is no longer a direct consequence.

	We will prove (conditional) convexity  in Corollary \ref{cor_prices_convex}. An explicit description of $\Pi_t(H)$ in terms of $\pi^{\text{sup}}_t(H)$ and $\pi^{\text{inf}}_t(H)$ will be given in Theorem \ref{thm_interval}.

	\begin{remark} 
		The essential supremum in the definition of $\pi^{\text{sup}}_t(H)$ can be taken in either $\cF_t$ or $\cF_T$. 
		As $\Pi_t(H) \subseteq L^0(\Omega, \cF_t, P)$
		they both coincide, since the essential supremum has a countable representation. 
	\end{remark}

	To achieve countable convexity of the set of equivalent martingale measures we use nonnegativity of the price process and triviality of the initial $\sigma$-algebra $\cF_0$ in the following lemma. As this result is standard, we omit its proof.
	\begin{lemma}
		\label{lem_emm_convex}
		$\cM_e$ is countably convex.
	\end{lemma}

	\begin{proposition}
		\label{prop_expectations_convex}
		For every $t \in \{0,\dots,T\}$, and for every $H\in L^0_+(\Omega, \cF_T, P)$ the set
		$$ \big\{ E_Q[H \mid \cF_t] : Q \in \cM_e \big\}$$
		is $\cF_t$-countably convex.
	\end{proposition}
	\begin{proof}
		Let $(Q^n)_{n \in \bN} \subseteq \cM_e$. By pasting, we may assume that all $Q^n$ agree on 
		$\cF_t$ without altering the set 
		$$ \big\{ E_{Q^n}[H \mid \cF_t] : n \in \bN \big\}.$$
		This 
		will be used throughout.
		
		Set $Q^* := \sum_n 2^{-n} Q^n$. By Lemma \ref{lem_emm_convex}, $Q^* \in \cM_e$.
		Denote by $Z^n := dQ^n/ d Q^*$ the associated densities. As $Q^* = Q^n$ on $\cF_t$ for each $n \in \bN$, we have 
		$$ Z^n_t = E_{Q^*}[Z^n \mid \cF_t] = 1. $$
		Since  $H \ge 0$, 
		monotone convergence implies for a sequence $(\lambda_t^n) \in L^0_+(\Omega, \cF_t, P)$ with $\sum_n \lambda_t = 1$, that
		\begin{align*}
			\sum_n \lambda^n_t E_{Q^n}[H \mid \cF_t] & = E_{Q^*}\Big[H \sum_n \lambda^n_t Z^n \mid \cF_t\Big]. 
		\end{align*}
		Set $Z := \sum_n \lambda^n_t Z^n > 0$. Note that
		$$ E_{Q^*}\Big[\sum_n \lambda^n_t Z^n \mid \cF_t\Big] = \sum_n \lambda^n_t = 1 $$ 
		and we may therefore define the measure $Q$ by
		$$ dQ / dQ^* := Z. $$
		Then,
		$$ E_{Q^*}\Big[H \sum_n \lambda^n_t Z^n \mid \cF_t\Big] = E_Q[H \mid \cF_t]. $$
		
		It remains to verify that $Q$ is indeed a martingale measure after $t$ (recall that  $Q$ and $Q^*$ agree on $\cF_t$).
		As the price process is nonnegative, its conditional expectation is well-defined, and we obtain by monotone convergence for $s \geq t$
		\begin{align*}
			E_Q[X_{s+1} \mid \cF_s] & = E_{Q^*}\Big[ \frac{Z}{Z_s} X_{s+1} \mid \cF_s\Big] \\
			& = \frac{1}{Z_s}  \sum_n \lambda^n_t E_{Q^*}[Z^n X_{s+1} \mid \cF_s]  \\
			& = \frac{1}{Z_s}  \sum_n \lambda^n_t Z^n_s E_{Q^n}[X_{s+1} \mid \cF_s] = X_s.
		\end{align*}
		This allows to conclude $Q \in \cM_e$.
	\end{proof}
	
	Note that, due to the integrability requirement, $\Pi_t(H)$ is not $\cF_t$-countably convex for every claim $H$. Even in the unconditional case this fails.

	\begin{corollary}Consider a claim $H \in L^0_+(\Omega, \cF_T, P)$.
		Further, let $(Q^n)_{n \in \bN} \subseteq \cM_e^H$ and \\$(\lambda_t^n)_{n \in \bN} \subseteq L^0_+(\Omega, \cF_t, P)$ with 
		$\sum_n \lambda_t^n = 1$.
		If $\sum_n \lambda_t^n E_{Q^n}[H \mid \cF_t]$ is finite-valued,
		it is contained in $\Pi_t(H)$.
	\end{corollary}
	\begin{proof}
		Due to Proposition \ref{prop_expectations_convex}, there exists $Q \in \cM_e$ with
		$$ \sum_n \lambda_t^n E_{Q^n}[H \mid \cF_t] = E_Q[H \mid \cF_t] \,.$$
		Now the claim follows from Lemma \ref{lem_price_finite}.
	\end{proof}
	
	\begin{corollary} Consider $t \in \{0,\dots,T\}$. Then,
		\label{cor_prices_convex}
		\begin{enumerate}[(i)]
			\item $\Pi_t(H)$ is $\cF_t$-convex for every claim $H\in L^0_+(\Omega, \cF_T, P)$, 
			
			\item $\Pi_t(H)$ is directed upwards for every claim $H\in L^0_+(\Omega, \cF_T, P)$, 
			
			\item  $\Pi_t(H)$ is $\cF_t$-countably convex for every bounded claim $H \in L^\infty(\Omega, \cF_T, P)$, and,
			
			\item for $H \in L^0_+(\Omega, \cF_T, P)$, any partition $(A^n)_{n \in \bN} \subseteq \cF_t$, and any sequence $(Q^n)_{n \in \bN} \subseteq \cM^H_e(P)$,
			$$ \sum_n \ind_{A^n} E_{Q^n}[H \mid \cF_t] \in \Pi_t(H) \,.$$
		\end{enumerate}
	\end{corollary}
	This observation  will imply that the no-arbitrage set $\Pi_t(H)$ is indeed an (of course random) interval for any $t \in \{0,\dots,T\}$. We will give a precise proof in Theorem \ref{thm_interval}.
	
	\begin{lemma}
		\label{lem_local_pi}
		For  $t \in \{0,\dots,T\}$,  $A \in \cF_t$ and  $H \in L^0_+(\Omega, \cF_T, P)$, it holds that
		$$ \Pi_t(\ind_A H) = \ind_A \Pi_t(H). $$
	\end{lemma}
	\begin{proof}
		Let $Q \in \cM_e$ such that $H \ind_A$ is integrable 
		with respect to $Q$.
		By Proposition \ref{prop_prices}, we may pick $\tilde{Q} \in \cM_e$
		such that $H \ind_{A^c}$ is integrable with respect to $\tilde{Q}$.
		By construction,
		\[
		E_Q[H \mid \cF_t] \ind_A + E_{\tilde{Q}}[H \mid \cF_t] \ind_{A^c}
		\]
		is finite-valued, and by Corollary \ref{cor_prices_convex} and Lemma \ref{lem_price_finite} there exists $Q^* \in \cM_e^H$ with
		$$ E_{Q^*}[H \mid \cF_t] = E_Q[H \mid \cF_t] \ind_A + E_{\tilde{Q}}[H \mid \cF_t] \ind_{A^c} $$
		and therefore
		\begin{align*} 
			E_{Q^*}[H \mid \cF_t] \ind_A &= E_Q[H\ind_A \mid \cF_t] \,. \qedhere \end{align*}
	\end{proof}

	\subsection{Characterization of the no-arbitrage bounds}
	In this section we show that for every claim $H$, the nonlinear expectation $$\bar{\cE}_t(H) = \esssup \{ E_Q[H \mid \cF_t] : Q \in \cM_e \}$$ can be computed by considering a subset of $\cM_e$ only: one can restrict to the set of martingale measures $\cM^H_e$ under which $H$ is integrable. 
	In particular, for every claim $H$, the nonlinear expectation $\bar{\cE}(H)$ agrees with the upper bound of the no-arbitrage interval $\pi_t^{\text{sup}}(H)$. This links the superhedging duality Proposition \ref{prop_superhedging_claims} with the pricing in financial markets.
	We start with two auxiliary results, before stating the mentioned result in Proposition \ref{prop_price_boundaries} below.

	\begin{lemma}
		\label{lem_exhaustion}
		Let $H\in L^0_+(\Omega, \cF_T, P)$. 
		Let $Y \in L^0(\Omega, \cF_t, P)$, and suppose $Y  < \bar{\cE}_t(H)$ on some $\cF_t$-measurable subset $A$ with $P(A) > 0$.
		Then, there exists a $\cF_t$-measurable set $B \subseteq A$ with $P(B) > 0$ and $\pi_t \in \Pi_t(H)$ such that
		$$ Y < \pi_t  \text{ on \(B\)} .$$

	\end{lemma}
	\begin{proof}
		As \(Y < \bar{\cE}_t(H)\) on \(A\), the definition of the essential supremum grants the existence of \(Q \in \cM_e\) and a \(\cF_t\)-measurable set $B \subseteq A$ with $P(B) > 0$ such that \( Y <  E_Q[H| \cF_t]\) on \(B\).
		Define \(C := \{E_Q[H| \cF_t] = \infty\} \cap B\). 
		If \(P(C) = 0\), Lemma \ref{lem_price_finite} together with Lemma \ref{lem_local_pi} implies that \(E_Q[H| \cF_t]\ind_B \in \Pi_t(H)\ind_B\). That is, there exists \(\pi_t \in \Pi_t(H)\) with
		\[
		\pi_t \ind_B = E_Q[H | \cF_t] \ind_B,
		\]
		and therefore 
		\[ Y < E_Q[H | \cF_t] = \pi_t \text{ on } B.\]
		Hence, we may assume \(P(C) > 0\). By monotone convergence, there exists 
		\(n \in \bN\) such that 
		\[
		P\big(Y \ind_C < E_Q[H \wedge n | \cF_t] \ind_C \big) > 0.
		\]
		We define, for \(k \in \{0, \dots, T\}\), \( X^{d+1}_k := E_Q[H \wedge n | \cF_k]\).
		Then, the extended market \((X, X^{d+1})\) is free of arbitrage and Proposition \ref{prop_prices} implies the existence of a martingale measure \(Q^*\) for the extended market such that \(H \in L^1(\Omega, \cF_T, Q^*)\).
		In particular, \(E_{Q^*}[H | \cF_t] \in \Pi_t(H)\) with
		\[
		E_{Q^*}[H | \cF_t] \geq E_{Q^*}[H \wedge n | \cF_t] = E_{Q^*}[ X^{d+1}_T | \cF_t] = X^{d+1}_t = E_{Q}[H \wedge n | \cF_t].
		\]
		Therefore,
		\[
		P\big(E_{Q^*}[H | \cF_t] \ind_C > Y \ind_C \big) \geq 
		P\big(E_{Q}[H \wedge n| \cF_t] \ind_C > Y \ind_C \big) > 0.
		\]
		Hence, \(\pi_t := E_{Q^*}[H | \cF_t]\) and \(\big\{E_{Q^*}[H | \cF_t] \ind_C > Y \ind_C \big\} \in \cF_t\) are as desired.
	\end{proof}
	
	\begin{proposition} \label{prop: exhaustion}
		Let $H\in L^0_+(\Omega, \cF_T, P)$.
		Let $Y \in L^0(\Omega, \cF_t, P)$, and suppose that $Y  < \bar{\cE}_t(H)$ on some $\cF_t$-measurable subset $A$.
		Then, there exists a $\cF_t$-measurable set $B \subseteq A$ with $P(B) = 
		P(A)$ and $\pi_t \in \Pi_t(H)$ such that
		$$ Y \leq \pi_t  \text{ on \(B\)} .$$
	\end{proposition}
	\begin{proof}
		If \(P(A) = 0\), set \(B:= \emptyset\). Then, \(Y \leq \pi_t\) on \(B\)
		for any \(\pi_t \in \Pi_t(H)\). Hence, we may assume \(P(A) > 0\).
		We use a Halmos-Savage argument.
		Set $\alpha := \sup\{ P(\{\pi_t \geq Y \} \cap A) : \pi_t \in \Pi_t(H) \}$. By Lemma \ref{lem_exhaustion}, $\alpha >0$.
		Pick a sequence $(\pi^n_t)_{n \in \bN} \subseteq \Pi_t(H)$ such that $P(\{\pi^n_t \geq Y \} \cap A)$ converges to $\alpha$.
		For $n \in \bN$, define $B^n := \{\pi^n_t \geq Y \} \setminus \cup_{k < n} \{\pi^k_t \geq Y \}$ and set
		$$ \pi_t := \sum_n \pi^n_t \ind_{B^n \cap A} + \pi^1_t \ind_{B^c \cup A^c},  $$
		where $B = \cup_n B^n$. 
		By Corollary \ref{cor_prices_convex}, $\pi_t \in \Pi_t(H)$.
		As $\{\pi_t \geq Y\} \cap A \supseteq \{\pi^n_t \geq Y\} \cap A $ for each $n \in \bN$, we have $P(\{\pi_t \geq Y\} \cap A) = \alpha$.
		It remains to show that $\alpha = P(A)$.
		Suppose $\{\pi_t < Y\} \cap A$ has positive probability.
		By Lemma \ref{lem_exhaustion}, there exists $\tilde{\pi}_t \in \Pi_t(H)$ with
		$$ \ind_C Y \leq \ind_C \tilde{\pi}_t$$
		for $C \subseteq \{ \pi_t < Y\} \cap A \subseteq B^c \cup A^c$ and \(P(C) > 0\).
		Setting
		$$ \hat{\pi}_t := \sum_n \pi^n_t \ind_{B^n \cap A} + \tilde{\pi}_t \ind_C + \pi^1_t \ind_{(B^c \cup A^c) \setminus C} $$
		yields an arbitrage-free price with $P(\{\hat{\pi}_t \geq  Y\} \cap A) > P(\{ \pi_t \geq Y\} \cap A) = \alpha$,
		a contradiction.
		Hence, \(\pi_t\) and \((\{\pi_t \geq  Y\} \cap A) \in \cF_t\) are as desired.
	\end{proof}

	\begin{proposition}
		\label{prop_price_boundaries}
		For every $H \in L^0_+(\Omega, \cF_T, P)$, we have the equalities
		\begin{align*}
			\esssup \{ E_Q[H \mid \cF_t] : Q \in \cM_e \} 
			& = \esssup \{ E_Q[H \mid \cF_t] : Q \in \cM_e^H \} 
		\end{align*}
		and
		\begin{align*}
			\essinf\{ E_Q[H \mid \cF_t] : Q \in \cM_e \} 
			& 	 = \essinf \{ E_Q[H \mid \cF_t] : Q \in \cM_e^H \} .
		\end{align*}
	\end{proposition}
	\begin{proof}
		Due to Proposition \ref{prop_prices}, it remains to show the first           equality.
		Clearly, \(\bar{\cE}_t(H) \geq \pi_t^{\text{sup}}(H)\).
		Set \( A := \{\bar{\cE}_t(H) = \infty \} \in \cF_t\).
		We start by showing
		\[
		\bar{\cE}_t(H) \ind_{A^c} = \pi_t^{\text{sup}}(H) \ind_{A^c}.
		\]
		To see this, note that for every \(Q \in \cM_e\), the expectation \(E_Q[H | \cF_t] \ind_{A^c}\) is finite-valued.
		Hence, by Lemma \ref{lem_price_finite}
		\[
		E_Q[H | \cF_t] \ind_{A^c} = E_Q[H \ind_{A^c} | \cF_t] \in \Pi_t(H\ind_{A^c}).
		\]
		As \(\Pi_t(H \ind_{A^c}) = \Pi_t(H)\ind_{A^c}\) by Lemma \ref{lem_local_pi}, there exists \(\pi_t \in \Pi_t(H)\)
		with
		\[
		E_Q[H|\cF_t]\ind_{A^c} = \pi_t \ind_{A^c}.
		\]
		We conclude that \(\bar{\cE}_t(H) \ind_{A^c} \leq \pi_t^{\text{sup}}(H) \ind_{A^c}\), and therefore 
		\( \bar{\cE}_t(H) \ind_{A^c} = \pi_t^{\text{sup}}(H) \ind_{A^c}\).
		Next, we prove
		\[
		\bar{\cE}_t(H) \ind_{A} = \pi_t^{\text{sup}}(H) \ind_{A}.
		\]
		To this end, let \(m \in \bN\). Then, using Proposition \ref{prop: exhaustion}, there exists \(\pi_t \in \Pi_t(H)\)
		such that \( m \leq \pi_t \) on a \(\cF_t\)-measurable subset \(B \subseteq A\) with \(P(B) = P(A)\).
		Since \(m \in \bN\) was arbitrary, we obtain 
		\[
		\pi_t^{\text{sup}}(H)\ind_A = \infty \ind_A = \bar{\cE}_t(H)\ind_A,
		\]       
		as desired.
	\end{proof}
	
	In view of Lemma \ref{lem_consistent_unique},
	Proposition \ref{prop_price_boundaries} is not an immediate consequence
	of the unconditional case, as the set $\cM_e^H$ is not stable.
	Moreover,  as the set $L^0_+(\Omega, \cF_T, P)$ is not symmetric, one has to show both equalities in Proposition \ref{prop_price_boundaries}.

	\section{A conditional version of the second fundamental theorem}
	\label{sec:FTAP}

	Equipped with efficient tools for nonlinear expectations we now prove a conditional
	version of the second fundamental theorem of asset pricing. We begin by studying those
	contingent claims which are attainable, i.e., claims which can be perfectly replicated by a hedging strategy.

	\subsection{Attainability of claims}\label{sec:attainability}
	Recall that a European contingent claim $H\in L^0_+(\Omega, \cF_T, P)$ is called \emph{attainable} (replicable) at time $t$ if there exists $\xi \in \pred$ and $H_t \in L^0(\Omega, \cF_t, P)$ such that
	$H = H_t + \rmG_t(\xi) $. 
	
	\begin{remark} Some immediate observations are due.
		\begin{enumerate}[(i)]
			\item If $H \in L^0_+(\Omega, \cF_T, P)$ is attainable at time $s$, then $H$ is attainable at time $t$ for every $t \geq s$.
			
			\item The converse of part (i) is not necessarily true. Using time consistency of $\pi^{\text{sup}}$, established in Proposition \ref{prop_superhedging_claims} and Proposition \ref{prop_price_boundaries}, we have
			$$ \pi^{\text{sup}}_s(H) = \pi^{\text{sup}}_s(\pi^{\text{sup}}_t(H)) $$
			and 
			$$ \pi^{\text{inf}}_s(H) = \pi^{\text{inf}}_s(\pi^{\text{inf}}_t(H)) .$$
			By Theorem \ref{thm_attainable} below, $H$ attainable at time $t$ is attainable at time $s \le t$ if and only if 
			$\pi^{\text{sup}}_t(H) = \pi^{\text{inf}}_t(H)$ is attainable at time $s$. \qedhere
		\end{enumerate}
	\end{remark}
	
	Recall that  $H \in L^0(\Omega, \cF_T, P)$ is called \emph{symmetric} with respect to $\cE_t$, if
	$$ \cE_t(H) = - \cE_t(-H). $$
	The next result gives a precise characterization of symmetry with respect to \(\cE_t\).
	
	\begin{lemma}
		\label{lem_symmetric_replicable}
		Let $H\in L^0_+(\Omega, \cF_T, P)$.
		Then, $H$ is symmetric with respect to $\cE_t$ if and only if $H$ is attainable at time $t$.
	\end{lemma}
	
	\begin{proof}
		Recall that by definition \eqref{defin:E*}, $\cE_t^*( \cdot ) = - \cE_t(- \cdot)$ and hence, as already established in Equation \eqref{def:E*},
		$$ \cE^*_t(H) = \esssup \{ H_t \in L^0(\Omega, \cF_t, P) : \exists \xi \in \pred : H_t + \rmG_t(\xi) \leq H \} $$
		is the largest subhedging  price. Due to Lemma \ref{lem_subhedging_price}, $\cE^*_t(H)$ is itself a subhedging  price.
		
		Now, suppose first that $H$ is attainable at time $t$, i.e., $H = H_t + \rmG_t(\xi)$
		for some $\xi \in \pred$. By Remark \ref{rem_orthogonality}, we obtain
		\[ \cE^*_t(H) = \cE^*_t(H_t) + \cE^*_t(G_t(\xi))  = H_t =
		\cE_t(H_t + G_t(\xi)) =  \cE_t(H)  .
		\]
		Therefore, \(H\) is symmetric.
		Second, let $\xi, \eta \in \pred$ such that
		\begin{align} \label{temp56}
			\cE^*_t(H) + \rmG_t(\xi) \leq H \leq \cE_t(H) + \rmG_t(\eta) \,.\end{align}
		If $H$ is symmetric, then $\cE_t(H) = \cE^*_t(H)$ is finite-valued and this implies
		$$ 0 \leq H - \cE_t(H) - \rmG_t(\xi) \leq \rmG_t(\eta - \xi) \,.$$
		The no-arbitrage assumption yields $G_t(\eta)=G_t(\xi)$. Since  $\cE_t(H) = \cE^*_t(H)$, Equation \eqref{temp56} yields that 
		$ H = \cE_t(H) + \rmG_t(H)$ and therefore $H$ is attainable.
	\end{proof}
	
	\begin{corollary}
		\label{cor_replicable_unique_price}
		Consider $H\in L^0_+(\Omega, \cF_T, P)$.
		$H$ is attainable at time $t$ if and only if
		\(E_{Q}[H \mid \cF_t]\) is constant over \(Q \in \cM_e\).
	\end{corollary}
	
	\begin{proof}
		Suppose first that \(H \in L^0_+(\Omega, \cF_T, P)\) is attainable at time \(t\). In particular, \(\cE_t(H)\) is finite-valued.
		Lemma \ref{lem_symmetric_replicable} shows that \(H\)
		is symmetric with respect to \(\cE_t\), and the dualities 
		established in Proposition \ref{prop_superhedging_claims} and 
		Proposition \ref{prop_subhedging_claims} yield that \(H\) is symmetric with respect to $\bar{\cE}_t$. Now,
		$(\bar{\cE}_t)^*(H) = \essinf \{ E_Q[H \mid \cF_t] : Q \in \cM_e \}$
		and symmetry $(\bar{\cE}_t)^*(H) = \bar{\cE}_t(H)$ is therefore equivalent to the constancy of $ E_Q[H \mid \cF_t]$ over $Q \in \cM_e$.
		Conversely, suppose that $ E_Q[H \mid \cF_t]$ is constant over $Q \in \cM_e$.
		Then, \(\bar{\cE}_t(H) = (\bar{\cE}_t)^*(H)\) is finite-valued, and the superhedging duality Proposition \ref{prop_superhedging_claims} implies that
		\(\cE_t(H)\) is finite-valued, too. Hence, we may apply the subhedging duality Proposition \ref{prop_subhedging_claims} again to obtain
		\[
		\cE_t(H) = \bar{\cE}_t(H) = (\bar{\cE}_t)^*(H) = \cE^*_t(H),
		\]
		i.e., symmetry of \(H\) with respect to \(\cE_t\). Thanks to Lemma \ref{lem_symmetric_replicable}, \(H\) is attainable at time \(t\).
	\end{proof}
	
	In view of Proposition \ref{prop_prices},
	Corollary \ref{cor_replicable_unique_price} extends the well-known result, that a claim is attainable if and only if it has a unique arbitrage-free price, to a conditional setting. See, e.g., Theorem 5.32 in \cite{FS} for the classical case.

	\begin{theorem}
		\label{thm_attainable}
		Let $H \in L_+^0(\Omega, \cF_T, P)$ be a European contingent claim.
		Then,
		\begin{enumerate}
			\item[(i)] $H$ is attainable at time $t$ if and only if 
			$$ \pi^{\text{inf}}_t(H) = \pi^{\text{sup}}_t(H),$$  i.e., when there is a unique arbitrage-free price at time $t$.
			
			\item[(ii)] If $H$ is not attainable at time $t$, then \( \pi^{\text{sup}}_t(H)\) is not an arbitrage-free price at time $t$.
			
			\item[(iii)] If $H$ is not attainable at time $t$, and \(\cE_t(H)\) is finite-valued,  then \( \pi^{\text{inf}}_t(H)\) is not an arbitrage-free price at time $t$.
		\end{enumerate}
	\end{theorem}

	\begin{proof}
		The first part, as just remarked, follows from Corollary \ref{cor_replicable_unique_price} in conjunction with Proposition \ref{prop_price_boundaries}. 
		Regarding part \((ii)\), the superhedging duality Proposition \ref{prop_superhedging_claims} grants the existence of \(\xi \in \pred\) with
		\[
		\pi^{\text{sup}}_t(H) + \rmG_t(\xi) \geq H, \text{ and } P(\{ \pi^{\text{sup}}_t(H) + \rmG_t(\xi) > H\}) > 0.
		\]
		If \(\pi^{\text{sup}}_t(H)\) is an arbitrage-free price, then \(\pi^{\text{sup}}_t(H)\) is necessarily finite-valued. Hence, we may assume \(\pi^{\text{sup}}_t(H) < \infty\).
		Let \(Q \in \cM_e\). Then, Lemma \ref{lem: integrability} implies
		\[
		\pi^{\text{sup}}_t(H) \geq E_Q[H | \cF_t], \text{ and } P(\{ \pi^{\text{sup}}_t(H) > E_Q[H | \cF_t]\}) > 0.
		\]
		As \(Q \in \cM_e\) was arbitrary, we conclude that \(\pi^{\text{sup}}_t(H) \not \in \Pi_t(H)\).
		To show \((iii)\), we argue similarly. As \(\cE_t(H)\) is finite-valued, the subhedging duality Proposition \ref{prop_subhedging_claims} grants the existence of \(\eta \in \pred\) with
		\[
		\pi^{\text{inf}}_t(H) + \rmG_t(\eta) \leq H, \text{ and } P(\{ \pi^{\text{inf}}_t(H) + \rmG_t(\eta) < H\}) > 0.
		\]
		Let \(Q \in \cM_e\). Then, Lemma \ref{lem: integrability} implies
		\[
		\pi^{\text{inf}}_t(H) \leq E_Q[H | \cF_t], \text{ and } P(\{ \pi^{\text{inf}}_t(H) < E_Q[H | \cF_t]\}) > 0.
		\]
		As \(Q \in \cM_e\) was arbitrary, we conclude that \(\pi^{\text{inf}}_t(H) \not \in \Pi_t(H)\).
	\end{proof}

	The next theorem  gives an explicit description of $\Pi_t(H)$ in terms of $\pi^{\text{sup}}_t(H)$ and $\pi^{\text{inf}}_t(H)$. In particular, it shows that the set of arbitrage-free prices is indeed a random interval with boundaries $\pi^{\text{sup}}_t(H)$ and $\pi^{\text{inf}}_t(H)$.
	
	\begin{theorem}	\label{thm_interval}
		Let $H \in L_+^0(\Omega, \cF_T, P)$ be a European contingent claim and suppose that  \(\pi^{\text{sup}}_t(H)\) is finite-valued. Then, the set of arbitrage-free prices $\Pi_t(H)$ is an $\cF_t$-measurable random interval, i.e.,
		$$ \Pi_t(H) = \{ \lambda_t \pi^{\text{sup}}_t(H) + (1-\lambda_t) \pi^{\text{inf}}_t(H) : \lambda_t \in L^0(\Omega, \cF_t, P),~ 0 < \lambda_t < 1 \text{ \(P\)-a.s.} \}. $$
	\end{theorem}
	
	In the case where $H$ is attainable, the (random) interval of arbitrage-free prices hence collapses to the singleton $\{ E_Q[H |\cF_t]\} $ with any $Q \in \cM_e^H$. When $H$ is not attainable, this interval is given by $(\pi^{\text{inf}}_t, \pi^{\text{sup}}_t)$. 
	
	\begin{proof}
		We start with the inclusion 
		$$ \Pi_t(H) \subseteq \{ \lambda_t \pi^{\text{sup}}_t(H) + (1-\lambda_t) \pi^{\text{inf}}_t(H) : \lambda_t \in L^0(\Omega, \cF_t, P),~ 0 < \lambda_t < 1 \text{ \(P\)-a.s.}\}. $$
		Let $\pi_t \in \Pi_t(H)$. Theorem \ref{thm_attainable} implies together with Lemma \ref{lem_local_pi} the equalities
		\begin{align*}
			\{ \pi_t = \pi^{\text{sup}}_t(H) \} & = \{ \pi^{\text{inf}}_t(H) = \pi_t = \pi^{\text{sup}}_t(H) \}, \\
			\{ \pi_t < \pi^{\text{sup}}_t(H) \} & = \{ \pi^{\text{inf}}_t(H) < \pi_t < \pi^{\text{sup}}_t(H) \}.
		\end{align*}
		Thus, we find $P(0 < \lambda_t < 1) = 1$, where
		$$ \lambda_t := \frac{\pi_t - \pi^{\text{inf}}_t(H)}{\pi^{\text{sup}}_t(H) - \pi^{\text{inf}}_t(H)} \ind_{\{\pi^{\text{sup}}_t(H) > \pi_t\}} + \delta \ind_{\{\pi^{\text{sup}}_t(H) = \pi_t\}} $$
		for any $\delta \in (0,1)$.
		By construction,
		$$ \pi_t = \lambda_t \pi^{\text{sup}}_t(H) + (1-\lambda_t) \pi^{\text{inf}}_t(H) \,.$$	
		
		Concerning the other inclusion, we may assume by Theorem \ref{thm_attainable} that $H$ is not attainable. Using Lemma \ref{lem_local_pi} and Corollary \ref{cor_prices_convex}, we may even assume that
		$ \pi^{\text{inf}}_t(H) < \pi^{\text{sup}}_t(H)$, see also the remark following this proof.
		Next, let \(Y \in L^0(\Omega, \cF_t, P)\) with $\pi^{\text{inf}}_t(H) < Y < \pi^{\text{sup}}_t(H)$. By Proposition \ref{prop: exhaustion}, there exists $\pi_t \in \Pi_t(H)$ such that $Y \leq \pi_t$.
		Similarly, there exists  $\tilde{\pi}_t \in \Pi_t(H)$ with $ \tilde{\pi}_t \leq Y$. Hence, conditional convexity of $\Pi_t(H)$ implies $Y \in \Pi_t(H)$.
	\end{proof}

	We end this section with some clarifying remarks regarding attainability of claims in the conditional setting.
	Let $t \in \{0, \dots, T\}$.
	Note that every claim $H$ decomposes as $H = H^A + H^B$, where
	$H^A$ is attainable at time $t$, and $H^B$ captures the non-attainability of $H$ in the sense that
	$$ \{\pi^{\text{inf}}_t(H^B) = \pi^{\text{sup}}_t(H^B) \} = \{\pi^{\text{inf}}_t(H) < \pi^{\text{sup}}_t(H) \} \,. $$
	Indeed, let $ A := \{\pi^{\text{inf}}_t(H) = \pi^{\text{sup}}_t(H) \}$ and set
	$$ H^A := H \ind_{A} \text{ and } H^B := H \ind_{A^c} \,.$$
	By virtue of locality,
	$$ \pi^{\text{inf}}_t(H^A) = \ind_A  \pi^{\text{inf}}_t(H) = \ind_A  \pi^{\text{sup}}_t(H) = \pi^{\text{sup}}_t(H^A) \,,$$
	whence $H^A$ is attainable at time $t$ by Theorem \ref{thm_attainable}.
	Using this decomposition, and Theorem \ref{thm_attainable}, one ends up with a tool for easy detection of prices yielding arbitrage, as captured
	in the following corollary.
	
	\begin{corollary}
		Let $H \in L^0_+(\Omega, \cF_T, P)$ be a European contingent  claim, and let $\pi_t$ be an $\cF_t$-measurable random variable.
		\begin{enumerate}[(i)]
			\item If $\{\pi_t > \pi^{\text{sup}}_t(H) \}$ has positive probability, then $\pi_t \notin \Pi_t(H)$.
			
			\item If $\{ \pi_t = \pi^{\text{sup}}_t(H) \} \cap \{ \pi^{\text{inf}}_t(H) < \pi^{\text{sup}}_t(H) \}$
			has positive probability, then $\pi_t \notin \Pi_t(H)$.
		\end{enumerate}
	\end{corollary}

	\subsection{Complete markets}
	We now establish a conditional version of the second fundamental theorem of asset pricing. To do so, we introduce the notation
	$$ \cM_e \odot_t Q^* := \{ Q \odot_t Q^* : Q \in \cM_e  \}, $$
	where $Q \odot_t Q^*$ denotes the pasting of $Q$ and $Q^*$ at time $t$ as defined in \eqref{def:pasting}.

	\begin{theorem}
		The following statements are equivalent: \label{FTAP2}
		\begin{enumerate}[(i)]
			\item The market is complete at time $t$, i.e., every European contingent claim $H \in L_+^0(\Omega, \cF_T, P)$ is attainable at time $t$.
			
			\item  Every European contingent claim $H \in L_+^0(\Omega, \cF_T, P)$ has a unique price at time $t$.
			
			\item For all \(Q, Q^* \in \cM_e\), the equality \(Q \odot_t Q^* = Q\) holds.
			
			\item For every $Q^* \in \cM_e$ the equality
			$ \cM_e = \cM_e \odot_t Q^*$ holds.
			
			\item There exists $Q^* \in \cM_e$ such that
			$ \cM_e = \cM_e \odot_t Q^* $.

		\end{enumerate}
	\end{theorem}
	
	\begin{proof}
		Theorem \ref{thm_attainable} provides the equivalence between market completeness at time $t$, and uniqueness of arbitrage-free prices at time $t$.
		
		As already mentioned in Remark \ref{rem_pasting}, the set of equivalent martingale measures is stable. This implies the inclusion $\cM_e \odot_t Q^* \subseteq \cM_e$ for every $Q^* \in \cM_e$.
		
		Now, let us first assume that every European contingent claim $H \in L_+^0(\Omega, \cF_T, P)$ has a unique price at time $t$, and let $Q, Q^* \in \cM_e$.
		We show the equality \(\tilde{Q} := Q \odot_t Q^* = Q\).   
		Using Remark \ref{rem_pasting}, we compute, for every $A \in \cF_T$,
		\begin{align*}
			E_{\tilde{Q}}[\ind_A] & = E_Q\big[ E_{Q^*}[\ind_A \mid \cF_t]\big]. 
		\end{align*}
		By assumption, the claim $\ind_A \in L^0_+(\Omega, \cF_T, P)$ has a unique arbitrage-free price at time $t$. In particular, the prices $E_{Q^*}[\ind_A \mid \cF_t]$
		and $E_Q[\ind_A \mid \cF_t]$ do agree. Hence,
		\[
		E_{\tilde{Q}}[\ind_A] = E_Q\big[ E_Q[\ind_A \mid \cF_t]\big]  = E_Q[\ind_A],
		\]
		and thus \(Q = \tilde{Q}\), as desired.
		Next, suppose that \((iii)\) holds, and let \(Q^* \in \cM_e\) be given.
		For \(Q \in \cM_e\), the equality \(Q = Q \odot_t Q^*\) shows \(Q \in \cM_e \odot_t Q^*\), and therefore \(\cM_e \subseteq \cM_e \odot_t Q^*\). We conclude that \((iv)\) holds.
		To finish the proof, suppose that the equality $\cM_e = \cM_e \odot_t Q^*$ holds for some $Q^* \in \cM_e$. We will argue that 
		every European contingent claim $H \in L_+^0(\Omega, \cF_T, P)$ has a unique price at time $t$, i.e., that $ \cM_e \ni Q \mapsto E_Q[H \mid \cF_t]$ is constant.
		To this end, let $\tilde{Q} \in \cM_e$. By assumption, there exists $Q \in \cM_e$ such that  $\tilde{Q} = Q \odot_t Q^*$.
		Using Remark \ref{rem_pasting} again, we obtain, for every  $H \in L_+^0(\Omega, \cF_T, P)$,
		$$ E_{\tilde{Q}}[ H | \cF_t] = E_{Q^*}[ H | \cF_t], $$
		and thus $\Pi_t(H)$ is indeed a singleton.	
	\end{proof}

	\begin{remark} ~
		\begin{enumerate}[(i)]
			\item Note that $ \cM_e = \cM_e \odot_T Q^*$ trivially holds, as 
			$ Q^1 \odot_T Q^2 = Q^1$ for any two measures $Q^1, Q^2$.
			This is in line with the observation that every claim $H \in L^0_+(\Omega, \cF_T, P)$ is attainable at time $T$.
			
			\item As $ Q^1 \odot_0 Q^2 = Q^2$ for any two measures $Q^1, Q^2$, we recover the classical version of the second fundamental theorem of asset pricing: market completeness is equivalent to the existence of exactly one equivalent martingale measure.
		\end{enumerate}
	\end{remark}

	Recall that, for a probability space \( (\Omega', \cG, Q)\), an atom is a set \(A \in \cG\) with \(Q(A) > 0\) such that for  every \(B \in \cG\) with \(B \subseteq A\) it holds that \(Q(B) \in \{0, Q(A)\}\).
	It is well-known that, in discrete time, the underlying probability space of a complete market wih \(d+1\) assets is purely atomic, and the number of atoms is bounded by \((d+1)^T\) (see, e.g., Theorem 5.37 in \cite{FS} or Theorem 6 in \cite{jacod1998local}).
	The following theorem gives the corresponding result in the conditional case
	and shows that, conditional on \(\cF_t\), the probability space is atomic.
	
	\begin{theorem} \label{thm: atomic}
		Let \(t \in \{0, \dots, T\}\). Suppose every European contingent claim $H \in L_+^0(\Omega, \cF_T, P)$ is attainable at time $t$. Then,
		\[
		\dim L^0(\Omega, \cF_T, P(\, \cdot \, | A) ) \leq (d+1)^{T-t}
		\]
		for every atom \(A\) of \(\cF_t\).
	\end{theorem}
	\begin{proof}
		Let \(A\) be an atom of \(\cF_t\). Then, \(\cF_t\) is \( P(\, \cdot \, | A) \)-trivial in the sense that \( P(B | A) \in \{0, 1\}\) for every \(B \in \cF_t\).
		We show that on the probability space \((\Omega, \cF_T, P(\, \cdot \, | A))\), the market with price process \( (X_s)_{s \in \{t,\dots,T\}}\) is complete:
		indeed, for \(H \in L^0_+(\Omega, \cF_T, P)\), there exists by assumption \(H_t \in L^0_+(\Omega, \cF_t,P)\) and \(\xi \in \pred\) such that
		\begin{equation} \label{eq: pf atoms}
			H = H_t + \rmG_t(\xi) \text{ \(P\)-a.s.}
		\end{equation}
		As \(P( \, \cdot \, | A) \ll P\), Equation \eqref{eq: pf atoms} holds \(P(\, \cdot \, | A)\)-a.s., too.
		Since \(H_t \in \cF_t\) is \(P(\, \cdot \, | A)\)-a.s. constant, completeness follows.
		
		By \cite[Theorem 5.37]{FS}, we obtain \(\dim L^0(\Omega, \cF_T, P(\cdot | A) ) \leq (d+1)^{T-t}\).
	\end{proof}

	\begin{appendix}
		\section{Nonlinear Expectations} \label{sec:nonlinear expectations}

		A central tool in our work are upper (and lower) bounds of the no-arbitrage interval. These are time-consistent, dynamic nonlinear expectations and in this section we present the required technical results on conditional nonlinear expectations together with related properties.
		
		Let $T \in \bN$ denote the final time horizon  and let $(\Omega, \cF)$ be a measurable space with a filtration $\bF= (\cF_t)_{t \in \{0,...,T\}}$. We assume $\cF_T = \cF$ and that $\cF_0$ is trivial. 
		
		A $P$-null set $A\subseteq \Omega$ is a possibly not measurable set being a subset of a measurable set $A'\in \cF$ with $P(A')=0$.
		Consider a set of probability measures \(\cP\) on \((\Omega, \cF)\).
		A set $A \subseteq \Omega$ is called a $\cP$\emph{-polar set}, if $A$ is a $P$-null set for every	$P \in \cP$. We denote the collection of $\cP$-polar sets by $\pol(\cP)$. 
		We say a property holds $\cP$-quasi surely, in short \(\cP\)-q.s., if it holds outside a $\cP$-polar set. If $\cP = \{P\}$, we write short $\pol(P)$ instead of $\pol(\{P\})$.
		
		For two subsets of probability measures \(\cP\) and $\cQ$, we call  $\cQ$  \emph{absolutely continuous} with respect to  $\cP$, denoted by $ \cQ \ll \cP$, if $\pol(\cP) \subseteq \pol(\cQ)$. We write $\cQ \sim \cP$, if $\cQ \ll \cP$ and $\cP \ll \cQ$.

		\bigskip
		
		On
		\( \sL^0(\Omega, \cF) = \{ X: \Omega \to \bR \colon X \ \cF\text{-measurable} \}\)
		we introduce the equivalence relation $\sim_\cP$ by
		$ X \sim_\cP Y \text{ if and only if } X = Y ~\cP \text{-q.s.} $.
		Then, we set
		$L^0(\Omega, \cF, \cP) := \sL^0(\Omega, \cF) / \cP$, and for $p\in [1, \infty)$
		\[
		L^p(\Omega, \cF, \cP) := \{X \in L^0(\Omega, \cF, \cP) : \sup_{P \in \cP} E_P[\abs{X}^p] < \infty\}.
		\]
		Next, define
		\[
		L^\infty(\Omega, \cF, \cP) := \{X \in L^0(\Omega, \cF, \cP) : \exists C > 0 : \abs{X} \leq C \ \cP \text{-q.s.}\}.
		\]
		
		With this notation, Proposition 14 in \cite{denis2011function} shows that for each $p \in [1,\infty]$, $L^p(\Omega, \cF, \cP)$ is a Banach space.
		
		The space $L^0(\Omega, \cF, \cP)$ can be equipped with a metric by
		$$ d \colon L^0(\Omega, \cF, \cP) \times L^0(\Omega, \cF, \cP) \to \bR_+, ~(X,Y) \mapsto \sup_{P \in \cP} E_P[ \abs{X-Y} \wedge 1 ] $$
		and this metric describes robust convergence in probability.

		We consider a set  $\cH \subseteq L^0(\Omega, \cF, \cP)$ containing all constants and set, for $t \in \{0,...,T\}$,
		$$ \cH_t := \cH \cap L^0(\Omega, \cF_t, \cP). $$
		
		The following definition introduces the notion of a conditional nonlinear expectation and the associated notion of a nonlinear dynamic expectation
		which is a set of conditional nonlinear expectations. 
		
		\begin{definition}
			We call a mapping $\cE_t : \cH \to \cH_t$ an  $\cF_t$\emph{-conditional nonlinear expectation}, if 
			\begin{enumerate}[(i)]
				\item $\cE_t$ is \emph{monotone}: for $X,Y \in \cH$ the condition $X \leq Y$ implies $\cE_t (X) \leq \cE_t (Y)$,
				\item $\cE_t$ \emph{preserves measurable functions}: for $X_t \in \cH_t$ we have $\cE_t(X_t) = X_t$.
			\end{enumerate} 
			We call $\cE = (\cE_t)_{t \in \{0,...,T\}}$ a \emph{nonlinear dynamic expectation}, if for every  $t \in \{0,...,T\}$ the mapping
			$\cE_t : \cH \to \cH_t$ is an $\cF_t$\emph-conditional nonlinear expectation.
		\end{definition}

		We introduce further properties which will be of interest in the context of dynamic nonlinear expectations. 
		First, we introduce some well-known properties regarding the set $\cH$, all in an appropriate conditional formulation. Denote $\cH_t^+:=\{X \in \cH_t: X \ge 0\}$. 
		We call $\cH$
		\begin{enumerate}[(i)]
			\item \emph{symmetric}, if \(-\cH = \cH\).
			\item \emph{additive}, if $\cH + \cH \subseteq \cH.$
			\item $\cF_t$\emph{-translation-invariant}, if $\cH + \cH_t \subseteq \cH$.
			\item $\cF_t$\emph{-convex}, if for $\lambda_t \in \cH_t$ with $0 \leq \lambda_t \leq 1$ we have
			$$ \lambda_t \cH + (1-\lambda_t) \cH \subseteq \cH. $$
			\item $\cF_t$\emph{-positively homogeneous}, if $ \cH_t^+ \cdot \cH \subseteq \cH$.
			\item $\cF_t$\emph{-local}, if $\ind_A \cH \subseteq \cH$ for every $A \in \cF_t$.
		\end{enumerate}
		Finally, we call $\cH$ \emph{translation-invariant}, if it is $\cF_t$-translation-invariant for every  $t \in \{0,...,T\}$ and do so in a similar fashion for the other properties.

		Next, we introduce well-known properties of nonlinear conditional expectations, all in an appropriate conditional formulation which are frequently used for example in the context of risk measures. An $\cF_t$-conditional expectation $\cE_t$ is called
		\begin{enumerate}[(i)]
			\item \emph{subadditive}, if $\cH$ is additive and $\cE_t(X+Y) \leq \cE_t(X) + \cE_t(Y)$ holds for $X,\  Y \in \cH$.
			
			\item $\cF_t$\emph{-translation-invariant}, if  $\cH$ is $\cF_t$-translation-invariant and  
			$$ \cE_t(X+X_t) = \cE_t(X) + X_t $$ 
			holds for any $X \in \cH$ and any $X_t \in \cH_t$.
			
			\item $\cF_t$\emph{-convex}, if $\cH$ is $\cF_t$-convex and for $\lambda_t \in \cH_t$ with $0 \leq \lambda_t \leq 1$ and $X, Y \in \cH$ the inequality
			$$ \cE_t\left(\lambda_t X + (1-\lambda_t) Y \right) \leq \lambda_t \cE_t(X) + (1-\lambda_t) \cE_t(Y) $$
			holds.
			
			\item $\cF_t$\emph{-positively homogeneous}, if $\cH$ is $\cF_t$-positively homogeneous and  
			$$\cE_t(X_t X ) = X_t \cE_t(X) $$
			holds for any $X \in \cH$ and any $X_t \in \cH_t^+$.
			
			\item $\cF_t$\emph{-sublinear}, if it is subadditive and $\cF_t$-positively homogeneous.
			
			\item $\cF_t$\emph{-local}, if $\cH$ is $\cF_t$ local and 
			$$ \cE_t(\ind_A X) = \ind_A \cE_t(X)$$ 
			for any $X \in \cH$ and any $A \in \cF_t$.
			
		\end{enumerate}                                          
		Finally, we call a dynamic expectation $\cE = (\cE_t)_{t \in \{0,...,T\}}$ translation-invariant, subadditive, convex or positively homogeneous,
		if for every $t \in \{0,...,T\}$ the $\cF_t$-conditional expectation $\cE_t$ has the corresponding property.
		
		\subsection{Sensitivity and time consistency}\label{sec:sensitivity and time consistency}
		In contrast to a classical expectation, a nonlinear expectation might contain only little information
		on underlying random variables. Sensitivity is a property which allows at least to separate zero from positive random variables:  we call an $\cF_t$-conditional nonlinear expectation $\cE_t$
		\emph{sensitive}, if for every $X \in \cH$ with $X \geq 0$ and $\cE_t(X) =0$ we have $X=0$.
		Similarly, we call the dynamic nonlinear expectation $\cE$ sensitive, if all $\cE_t$, $t =0,\dots,T$ are sensitive.

		\begin{remark}[Sensitivity in the context of risk measures]
			Recall that a conditional risk measure $\rho_t$ on $L^\infty(P)$ is called \emph{sensitive}, if for $A \in \cF$ with $P(A) >0$ and every $\delta > 0$ the set 
			$$ \{ \rho_t( - \delta \ind_A) > 0 \}$$
			has positive probability. Under the bijection $\rho_t \mapsto (X \mapsto \rho_t(-X))$ between risk measures and translation-invariant expectations, both notions coincide: indeed, if $\cH$ is local, then every conditional nonlinear expectation $\cE_t$ is sensitive if and only if for every $A \in \cF$ with  $A \notin \pol(\cP)$ and $\delta>0$ one has 
			\begin{align*} & \{ \cE_t(\delta \ind_A) > 0 \} \notin \pol(\cP) .
			\end{align*}
		\end{remark}

		\emph{Time consistency} is an important property in the context of dynamic risk measures, which has been intensively studied, see, e.g., \cite{acciaio2011dynamic}, \cite{artzner2007coherent}, \cite{bion-nadal2008}, \cite{cheridito2006}, \cite{delbaen2006}, \cite{detlefsen2005conditional}, \cite{riedel2004}  and the references therein.
		Let us introduce the appropriate definition for dynamic nonlinear expectations:
		we call a dynamic expectation $\cE$ \emph{time-consistent}, if for $s, t \in \{0,...,T\} $ with $s \leq t$ the equality
		$$ \cE_s = \cE_s \circ \cE_t $$
		holds.

		Since $\cF=\cF_T$, $\cE_T$ is the identity and hence,  every expectation is time-consistent between $T-1$ and $T$, i.e.,
		$$ \cE_{T-1} \circ \cE_T = \cE_{T-1}. $$
		
		\begin{remark}[Extension of time consistency to stopping times]
			For simplicity, we restrict our definition of time consistency to deterministic times $s, t \in \{0,\dots,T\}$. This can easily be generalized. Indeed, let $\tau$ be a stopping time with values in $\{0,\dots, T\}$. Given $(\cE_t)_t$, we define
			$$ \cE_\tau(H) := \sum_s \ind_{\{\tau = s\}} \cE_s(H) \,.$$
			If $(\cE_t)_t$ is time-consistent, then for any two such stopping times $\sigma, \tau$ with $\sigma \leq \tau$, the equality
			$$\cE_\sigma \circ \cE_\tau = \cE_\sigma $$
			holds whenever $\cE$ is translation-invariant and local.
		\end{remark}
		
		The remarkable connection between sensitivity and time consistency can already be seen from the simple observation that a time-consistent dynamic expectation is already sensitive, if $\cE_0$ is sensitive.

		\begin{remark}
			\label{rem_sensitivity_polars}
			Let  $\cP$ and $\cQ$ be two sets of probability measures on \((\Omega, \cF)\), and let \\
			$\cE_t : L^\infty(\Omega, \cF, \cQ) \to L^\infty(\Omega, \cF_t,\cQ)$ be a conditional nonlinear expectation. Then, $\cE_t$ is well-defined on $L^\infty(\Omega, \cF, \cP)$ if and only if $\cQ \ll \cP$.
			However, for $H \in L^\infty(\Omega, \cF, \cP)$ the evaluation $\cE_t(H)$ is a priori only an element of $L^\infty(\Omega, \cF_t, \cQ)$. For it to be well-defined in $L^\infty(\Omega, \cF, \cP)$ we require $\cQ \sim \cP$ on $\cF_t$. 
			Hence, if $\cQ \ll \cP$ and $\cQ \sim \cP$ on $\cF_t$, the conditional nonlinear expectation $\cE_t$ induces a nonlinear expectation
			$ \bar{\cE}_t : L^\infty(\Omega, \cF, \cP) \to L^\infty(\Omega, \cF_t, \cP)$.
			In case $\cE_t$ is sensitive, sensitivity of $\bar{\cE}_t$ is equivalent to
			$ \cP \sim \cQ$.		
		\end{remark}

		Lemma \ref{lem_acceptance_consistent} below generalizes the well-known result that the 
		acceptance sets of time-consistent expectations are decreasing:  if
		$\cE$ is time-consistent, then 
		$$  \{ \cE_s \leq 0 \}  \supseteq \{\cE_t \leq 0\} $$
		for $s \leq t$.
		See \cite[Lemma 11.11]{FS} for the corresponding risk measure result, and note that only preservation of constants is needed for the proof.

		\begin{lemma}
			\label{lem_acceptance_consistent}
			Let $\cE$ be a time-consistent, local nonlinear dynamic expectation and $t \in \{0,\dots,T\}$. Then,  for $H \in \cH$ the condition
			$\cE_t(H) \leq 0$ implies $\cE_s(\ind_A H) \le 0$ for all $A \in \cF_t$ and all $s \le t$.
			If $\cE_0$ is sensitive, then the converse is also true.	
		\end{lemma}
		\begin{proof}
			First, locality implies that $\cE_t(H \ind_A) = \ind_A \cE_t(H) \leq 0$. Together with monotonicity we obtain
			$$ \cE_s(\ind_A H) = \cE_s(\ind_A\cE_t(H)) \leq 0 .$$
			Now, suppose $\cE_0$ is sensitive. Then, $\cE_s$ is sensitive.
			To show that $\cE_t(H) \leq 0$, it thus suffices to show
			$$ \cE_s(\ind_A \cE_t(H)) = 0 $$
			for $A := \{ \cE_t(H) \geq 0\} \in \cF_t$. However, as above,
			$$ \cE_s(\ind_A \cE_t(H)) = \cE_s(\ind_A H) $$
			and the latter vanishes by assumption.	 
		\end{proof}

		Let $\cE^*_0$ be a $\cF_0$-conditional expectation. A \emph{dynamic extension} of $\cE^*_0$ is a dynamic expectation $\cE = (\cE_t)_{t \in \{0,...,T\}}$ such that $\cE_0 = \cE^*_0$. 
		The following Lemma clarifies and generalizes previous uniqueness results, such as \cite[Lemma 1]{cohen2012quasi} showing that every coherent expectation has at most one coherent and time-consistent dynamic extension.
		Here, uniqueness is  understood as uniqueness up to a polar-set, where
		the set of reference measures comes from the representation of $\cE_0$ in the sense of \cite[Theorem I.2.1.]{peng2010nonlinear}. 
		Regarding the mentioned result,
		note that for any collection $\cP$ of probability measures, the associated nonlinear expectation
		$\sup_{P \in \cP} E_P[\cdot]$ is sensitive; see Remark \ref{rem_sensitivity_polars}.
		
		Recall that \(\cP\) is \emph{dominated} if there exists a probability measure 
		\(P\) on \( (\Omega, \cF)\) with \(\cP \ll \{P\}\), i.e., every \(P\)-null set is \(\cP\)-polar. In this case, the Halmos-Savage Lemma guarantees the existence of a countable collection \(\{P_n \colon n \in \bN\} \subseteq \cP\) with
		\(\{P_n \colon n \in \bN\} \sim \cP\). In particular, there exists a measure \(P^*\) (not necessarily contained in \(\cP\)) such that \(\cP \sim P^*\).
		Consequently, for any set of random variables \(M \subseteq L^0(\Omega, \cF, \cP) = L^0(\Omega, \cF, P^*)\), there exists a random variable called the \(\cP\)-essential infimum and denoted by $\cP-\essinf M$ such that 
		\begin{enumerate}
			\item[(i)] \(\cP-\essinf M \leq Y\) \(\cP\)-q.s. for every \(Y \in M\), 
			
			\item[(ii)] \(\cP-\essinf M \geq Z\)  \(\cP\)-q.s. for every random variable \(Z \) satisfying \(Z \leq Y\) \(\cP\)-q.s. for every \(Y \in M\).
		\end{enumerate}

		If \(\cP\) is not dominated, the \(\cP\)-essential infimum might not exist, and it has in general no countable representation. See \cite{cohen2012quasi} and \cite{liebrich2020} for further details. In light of the financial applications we have in mind, we will assume in the next lemma that \(\cP\) is dominated.
		
		\begin{lemma}
			\label{lem_consistent_unique}
			Assume that \(\cP\) is dominated. Then, every sensitive $\cF_0$-conditional expectation $\cE_0$ 
			on a symmetric set \(\cH\)
			has at most one translation-invariant, local, time-consistent dynamic extension $(\cE_t)_t$. If it exists, it is given by
			$$ \cE_t(H) = \cP-\essinf \{H_t \in \cH_t : H-H_t \in A_t\}, $$
			where 
			$$ A_t := \{H \in \cH : \cE_0(\ind_A H) \leq 0, ~\forall A \in \cF_t \} \,. $$
			
		\end{lemma}
		\begin{proof}
			Lemma \ref{lem_acceptance_consistent} characterizes for $t \geq 1$ the acceptance set $A_t := \{H \in \cH \colon \cE_t \leq 0\}$ solely in terms of $\cE_0$ and $\bF$. Indeed, it yields that
			$$ A_t = \{H \in \cH : \cE_0(\ind_A H) \leq 0 ~ \forall A \in \cF_t \} \,. $$
			This allows to recover every translation-invariant nonlinear expectation 
			on a symmetric set from its acceptance set through the representation
			\begin{align*}
				\cE_t(H) & = \cP-\essinf \{H_t \in \cH_t : H_t \geq \cE_t(H) \} \\
				& = \cP-\essinf \{H_t \in \cH_t : H-H_t \in A_t\}.
			\end{align*}
			Summarizing, we have  not only shown uniqueness of the extension, but even obtained an explicit expression.
		\end{proof}

		Next, we verify that translation-invariance implies locality if $\cH \subseteq L^\infty(\Omega, \cF, \cP)$. This sharpens \cite[Proposition 2]{detlefsen2005conditional}, as it shows that every conditional risk measure on $L^\infty(\Omega, \cF, \cP)$ is local; convexity is not required.
		In particular, for every probability measure \(P\),
		every dynamic risk-measure on $L^\infty(\Omega, \cF, P)$ has at most one time-consistent extension. Moreover, one can show that not every coherent risk measure has a time-consistent extension. 
		
		\begin{proposition}
			\label{prop: trans}
			Every translation-invariant expectation on a local set $\cH \subseteq L^\infty(\Omega, \cF, \cP)$ is local.
		\end{proposition}
		\begin{proof}
			Let $\cE_t$ be translation-invariant and $A \in \cF_t$.
			Further, let \(H \in \cH\).
			The inequality
			$$ \ind_A H - \ind_{A^c} \norm{H}_\infty \leq  H  \leq \ind_A H + \ind_{A^c} \norm{H}_\infty,$$ 
			yields
			$$ \cE_t(H) \geq \cE_t(\ind_A H - \ind_{A^c} \norm{H}_\infty),$$
			and additionally
			$$ \cE_t(H) \leq \cE_t(\ind_A H + \ind_{A^c} \norm{H}_\infty).$$
			Multiplying both inequalities with $\ind_A$  gives
			$$ \ind_A \cE_t(H) = \ind_A \cE_t(\ind_A H), $$
			and thus
			\begin{align*}
				\cE_t(\ind_A H) &= \ind_A \cE_t(\ind_A H) + \ind_{A^c} \cE_t(\ind_A H) \\
				&= \ind_A \cE_t(H) + \ind_A \ind_{A^c} \cE_t(\ind_A H) \\
				&= \ind_A \cE_t(H). \qedhere
			\end{align*}
		\end{proof}

	\end{appendix}

\end{document}